\documentclass[sigplan,10pt]{acmart} 
\renewcommand\footnotetextcopyrightpermission[1]{}
\usepackage{amsmath,amsfonts,mathtools, mathrsfs}
\usepackage[toc, page, title]{appendix}
\usepackage{algpseudocode}
\usepackage{algorithmicx}
\usepackage{algorithm}
\algrenewcommand{\algorithmiccomment}[1]{\hskip3em// #1} 

\usepackage{tikz}
\usepackage{amsmath}
\usepackage{soul}
\usepackage{graphicx}
\usepackage{subcaption}
\usepackage{textcomp}
\usepackage{xcolor, colortbl }
\usepackage[normalem]{ulem}

\usepackage{balance}
\usepackage{enumerate}
\usepackage{etoolbox}
\usepackage{verbatim}
\usepackage{float}
\usepackage{amsthm} 
\usepackage{gensymb}
\usepackage{multirow}
\usepackage{xspace}
\usepackage{listings} 

\PassOptionsToPackage{hyphens}{url}
\usepackage{hyperref}

\newenvironment{smitemize}%
  {\begin{list}{$\bullet$}%
     {\setlength{\parsep}{0pt}%
      \setlength{\topsep}{0pt}%
      \setlength{\itemsep}{0pt}}}%
  {\end{list}}

\newbool{IsPrintComment}
  \booltrue{IsPrintComment}   
\newcommand{\lineCut}[1]
{
    \ifbool{IsPrintComment}
    {
        {\sout{#1}}
    }
    \  
}

\newcommand{\squishlist} 
{
    \begin{list}{$\bullet$}
    {
        \setlength{\itemsep}{0pt}      \setlength{\parsep}{3pt}
        \setlength{\topsep}{3pt}       \setlength{\partopsep}{0pt}
        \setlength{\leftmargin}{1.5em} \setlength{\labelwidth}{1em}
        \setlength{\labelsep}{0.5em}
    }
}

\newcommand{\squishend}
{
    \end{list}
}

\newcommand{\landon}[1]
{
    \ifbool{IsPrintComment}
    {
        {\bf \color{cyan} LC: #1}
    } 
    \  
}
\newcommand{\landonC}[1]
{
    \ifbool{IsPrintComment}
    {
        {\color{magenta} #1}
    }
    {
        { #1 }
    }
	\  
}

\newcommand{\shadi}[1]
{
    \ifbool{IsPrintComment}
    {
         {\color{orange}(SN) #1}
    }
	\  
}

\definecolor{darkgreen}{RGB}{1,150,33}
\newcommand{\shadiC}[1]
{
    \ifbool{IsPrintComment}
    {
         {\color{darkgreen} #1}
    }
	{
		{ #1}
	}
}
\newcommand{\joseph}[1]
{
    \ifbool{IsPrintComment}
    {
         {\color{blue}(JL) #1}
    }
    {
        { #1}
    }
	\  
}

\newcommand{\todo}[1]
{
    \ifbool{IsPrintComment}
    {
         {\color{red} [#1]}
    }
	\  
}

\newcommand{\needreview}[1]
{
    \ifbool{IsPrintComment}
    {
         {\color{blue} (New Edit): #1}
    }
    {
        { #1 }
    }
    \
}

\newcommand{\name}{BumbleBee}

\newcommand{\system}{BumbleBee\xspace}

\newcommand\unit{\,}
\lstdefinestyle{lua}{
  language=[5.1]Lua,
  keywordstyle=\color{magenta},
  stringstyle=\color{blue},
  commentstyle=\color{black!60},
  basicstyle=\ttfamily\footnotesize,
  captionpos=b,
  frame = single,
  numbers=left,
  stepnumber=1,
}
\let\oldbibitem\bibitem
\def\bibitem{\vfill\oldbibitem}

\setcopyright{none}

\begin{document}

    \date{}
    \title{\Large \bf \system: Application-aware adaptation for container orchestration}
    

    \author{HyunJong Lee}
    \affiliation{
      \institution{University of Michigan}
      \city{}
      \country{}
    }

    \author{Shadi Noghabi}
    \affiliation{
      \institution{Microsoft Research}
      \city{}
      \country{}
    }

    \author{Brian D. Noble}
    \affiliation{
      \institution{University of Michigan}
      \city{}
      \country{}
    }

    \author{Matthew Furlong}
    \affiliation{
      \institution{Microsoft}
      \city{}
      \country{}
    }

    \author{Landon P. Cox}
    \affiliation{
      \institution{Microsoft Research}
      \city{}
      \country{}
    }


    \begin{abstract}
        Modern applications have embraced separation of concerns as a first-order
organizing principle through the use of containers, container orchestration,
and service meshes. However, \emph{adaptation} to unexpected network variation
has not followed suit. We present \system, a lightweight extension to the
container ecosystem that supports application-aware adaptation. \system
provides a simple abstraction for making decisions about network data using
application semantics. Because this abstraction is placed within the
communications framework of a modern service mesh, it is closer to the point
at which changes are detected, providing more responsive and effective
adaptation than possible at endpoints. 

    \end{abstract}

    \maketitle
    \pagestyle{plain}

    \section{Introduction}
\label{sec:intro}



Containers~\cite{merkel-linux14} have become the dominant framework
for deploying large-scale applications~\cite{vmware2020}, with good
reason. Composing applications as a set of communicating
microservices~\cite{newman-2015} allows small teams to collaborate
across well-defined interfaces with minimal friction. Containers
simplify deployment of individual components, and these components can
be separately updated.

Container ecosystems also provide \emph{separation of concerns} as a
top-level goal. Orchestration frameworks~\cite{hightower-oreilly17}
allow declarative specifications of service goals, handling task
deployment, placement, scaling, and management. Service
meshes~\cite{sharma-2019} coordinate the communication between these
managed deployments, including health monitoring, routing, and load
balancing. 

Collectively, this architecture attempts to maintain consistent
performance and reliability while keeping resource usage proportional
to load. As conditions change, orchestrators can launch tasks to
satisfy bursts of new requests, kill tasks when utilization drops, and
load-balance traffic among tasks. However, such decisions are
necessarily imperfect, as it is impossible to predict changes in
availability of and demand for network resources connecting these
components.

Application-oblivious responses to network fluctuations, such as TCP
congestion control, try to provide fair allocation of limited
resources, but only the application knows how best to adjust its
\emph{fidelity} in response to changing network
conditions~\cite{noble-sosp97}. For example, a stream processing
workload might forgo less valuable computations rather than suffer
prohibitive latency when some component pathways become congested.
Likewise, a video streaming application can reduce its fidelity by
serving lower bitrate content when bandwidth becomes scarce.

Unfortunately, orchestration platforms do not provide clean
abstractions for application-aware adaptation. The
alternative---leaving adaptation to application endpoints---runs
counter to the goal of separation of concerns. To remedy this, we
present \emph{\system}, a platform that leverages common
infrastructure to provide application informed, in-network adaptation
via a simple set of abstractions.

\system{'s} design is based on a study of a variety of adaptive
applications. These applications collectively exhibit four common
adaptive patterns in response to changing network conditions:
\emph{dropping}, \emph{reordering}, \emph{redirecting}, or
\emph{transforming} messages. These patterns can be supported by
simple operations in the critical path of message processing,
augmented by asynchronous callbacks to other components. These
callbacks support both distributed monitoring on behalf of
applications as well as more computationally intensive tasks on the
messages themselves.

In \system, applications supply concise scripts that describe these
adaptive patterns. These scripts execute within
\emph{sidecars}~\cite{klein-srecon17}, which are user-level network
proxies that are an established part of the service mesh abstraction
for traffic monitoring and management. This small incremental change
to an existing mechanism allows application-aware adaptation to
benefit from the separation of concerns already provided by the
container ecosystem. Applications need not take on the task of network
monitoring, and adaptation strategies can be updated and repurposed
without inspecting complex sourcecode or deploying new
containers. Finally, this allows placement of adaptive mechanisms
closer to the point at which change occurs within the network,
improving agility over end-to-end approaches.

This paper makes the following contributions:
\begin{smitemize}
    \item We identify four common patterns applications use to adapt
      to network variability.
    \item We design and implement a single in-network abstraction to
      implement all of these patterns, informed by application needs.
    \item Experiments with our prototype show that (1) by using
      \system, ML applications at the edge can utilize cloud resources
      when available and operate without interruption when
      disconnected, (2) \system increases the number of deadlines met
      between 37.8\% and 23x on the Yahoo! stream-processing
      benchmark, (3) \system reduces stalled playback by 77\% during
      HLS video streaming under real-world network conditions, and (4)
      \system adds less than 10\% overhead to the 99th percentile
      request latency compared to a baseline sidecar.
\end{smitemize}


    \section{Background}
\label{sec:background}

\begin{figure}[]
	\centering
	\includegraphics[width=\columnwidth]{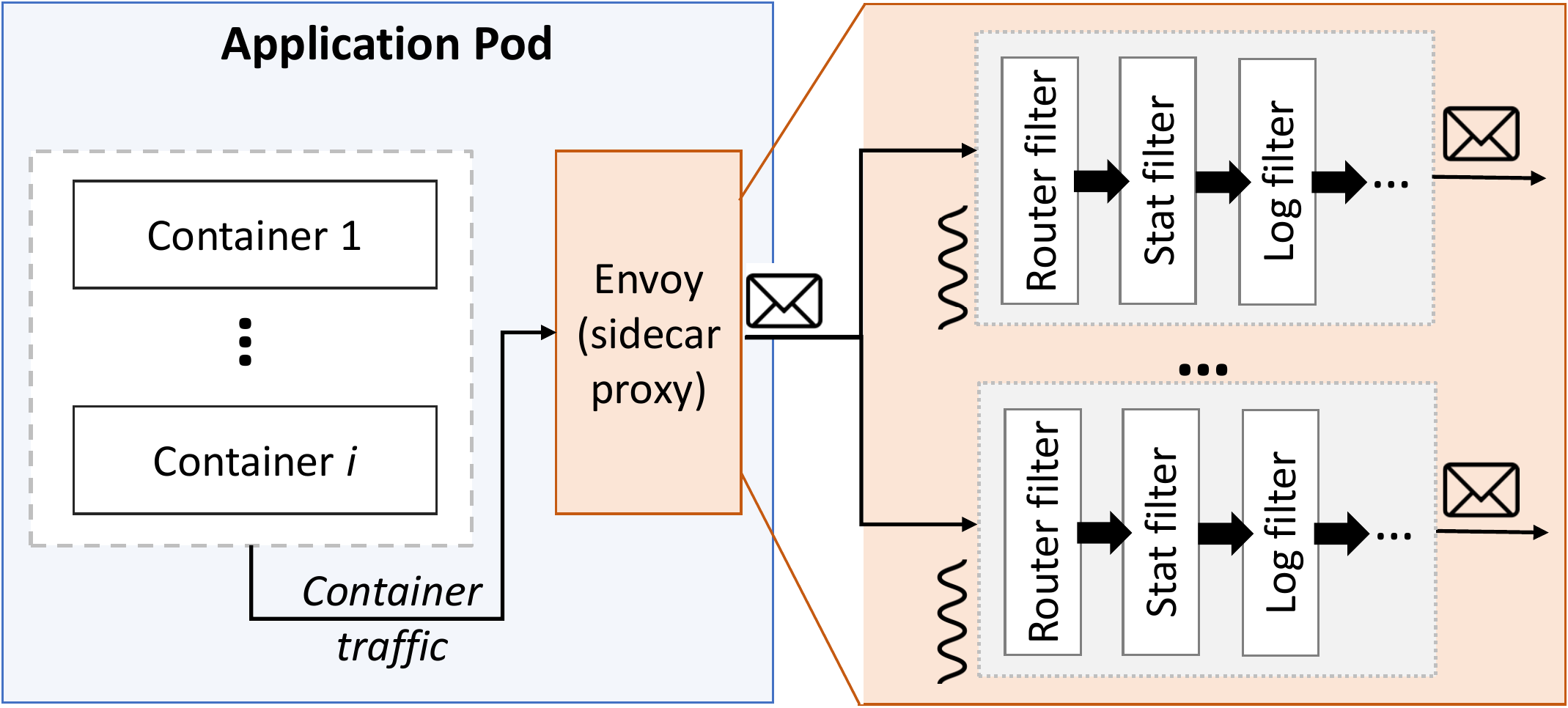}
	\caption{Envoy sidecars interpose on a pod's network communication.}
	\label{fig:envoy}
\end{figure}

Applications are increasingly written in a \emph{containerized
framework}~\cite{banga-osdi99,merkel-linux14,soltesz-eurosys07} as a
collection of communicating
\noindent\emph{microservices}~\cite{newman-2015}. These frameworks provide many
advantages: a strict decomposition of tasks, a consistent deployment
model allowing in-place updates, declarative capture and preservation
of system dependencies, and lightweight resource isolation and
monitoring. Such ecosystems explicitly provide for \emph{separation of
concerns} through architectural decisions. Application writers need
not be concerned with task creation, monitoring, placement, or
scaling, relying instead on \emph{container orchestration
frameworks}~\cite{hightower-oreilly17,verma-eurosys15}. Likewise, they
need not actively manage the communication between emplaced
tasks. Instead, a \emph{service mesh}~\cite{li-2019,sharma-2019}
provides reliable, fault-tolerant, load-balanced communication across
complex topologies of task deployment. This section describes these
frameworks, with an eye to \system's
integration with them.

\noindent{\bf Containers:} Docker~\cite{merkel-linux14} is a container-based
virtualization platform that provides process-level performance and
security isolation; such platforms have become the standard unit to
manage and deploy software in the cloud. Container images include all
of the user-level state required to launch an application, including
binaries, support libraries, and configuration. Each container
typically implements a single component microservice of the overall
application, providing an API to the other constituent components.

\noindent{\bf Container Orchestration:} Kubernetes~\cite{hightower-oreilly17}
automates deployment, scaling, and management of distributed,
containerized applications. The unit of deployment in Kubernetes is a
\emph{pod}. A pod is a set of containers that run under the same
operating system kernel and share the same underlying physical
resources, such as cores and disks. Because containers within a pod
share a machine they can communicate cheaply via local storage or
intra-kernel messaging.

Developers write configuration manifests describing 
how Kubernetes should deploy an application on a set of physical
or virtual machines, e.g., which container images to use, how
containers are grouped into pods, and which ports each pod needs. The
manifest also describes runtime goals for an
application, such as pod replication factors, load balancing among
replicas, and an auto-scaling policy.


\noindent{\bf Service Mesh:} Service meshes~\cite{li-2019} manage
inter-pod communications within Kubernetes. They provide service
discovery, peer health monitoring, routing, load balancing,
authentication, and authorization. This is done via the \emph{sidecar
pattern}~\cite{burns-hotcloud16}, in which a user-level network proxy
called Envoy~\cite{klein-srecon17} is transparently interposed between
each pod and its connection to the rest of the system; applications
are oblivious to the sidecar and its mechanisms. Each Envoy instance
is populated with {\tt iptable} rules to route incoming and outgoing
packets through the sidecar, as shown in Figure~\ref{fig:envoy}. This
architecture makes the Envoy sidecar an ideal place to implement
application-aware adaptation. It allows application writers to focus
only on the needs of adaptation as data traverses the network, without
having to integrate it with the application's behavior as prior
systems did~\cite{noble-sosp97,fox-asplos96}. We use the
Istio~\cite{sharma-2019} implementation in our prototype.

An Envoy sidecar has a pool of worker threads, mapped to the underlying
threads exposed to this container. Workers block on ingress/egress
sockets, and are invoked on a per-message basis. On invocation, the
Envoy worker passes the message through one or more application-specific
\emph{filters}. Filters are small, stateless
code snippets that operate on individual messages. Filters have full
access to a message and can perform simple operations, such as
redirection, dropping, and payload transformation.  Developers
commonly use Envoy filters for monitoring and traffic shaping, such as
collecting telemetry, load-balancing, and performing A/B testing.
Envoy supports filters at several layers of the network stack.



    \section{Design and implementation}
\label{sec:impl}
\label{sec:design}

In designing \system{}, we kept three goals in mind. First, we
followed the container ecosystem's core principle of \emph{separation
of concerns}, isolating the application's logic from adaptation
decisions. The existing infrastructure of orchestration frameworks and
service meshes made this particularly attractive. Second, we
\emph{kept interfaces as narrow as possible}. There were places where
\system{} needs some additional information or functionality, but
those were chosen only reluctantly. Third, we erred on the side of
\emph{simple and inexpensive} in designing the interface exposed by
\system{} whenever possible.

\begin{figure}[]
	\centering
	\includegraphics[width=\columnwidth]{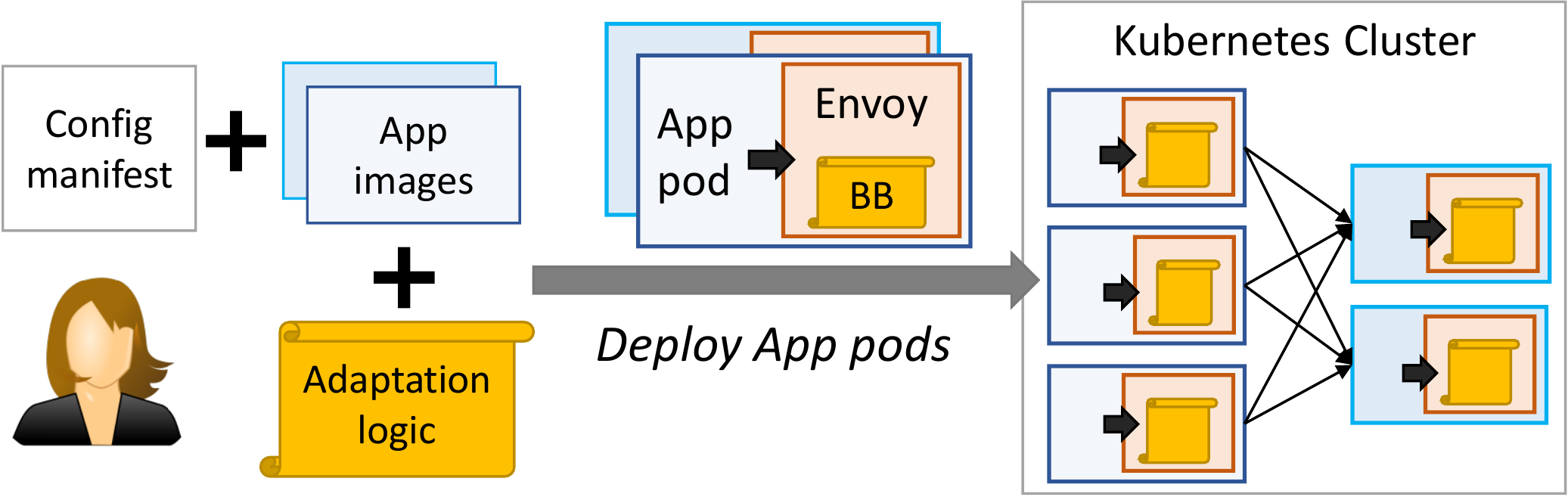}
	\caption{Kubernetes deployment of a distributed application with \name{} enabled.}
	\label{fig:k8-deployment}
\end{figure}

\system{'s} overall architecture is illustrated in
Figure~\ref{fig:k8-deployment}. Authors define applications through a
YAML manifest that describes the set of application container images
and their corresponding configurations. When deployed, the Istio
service mesh co-locates an Envoy sidecar proxy with each Kubernetes
pod; this sidecar is interposed on all traffic to and from the
pod. Applications supply \system{} adaptation logic as simple Lua
scripts~\cite{ierusalimschy-spe96}, deployed in the
sidecars as Envoy filters~\cite{sharma-2019}. All elements in a
container ecosystem can be updated in place; thus these scripts can be
changed on the fly without stopping or re-deploying the overall
application.

To determine how best to frame the abstractions provided by \system,
we surveyed a variety of existing adaptive systems, including video
streaming~\cite{yan-nsdi20, mao-sigocmm17, Wang05} and
conferencing~\cite{De08, Zhang12, Robinson16, Jiang16}, sensor
networks~\cite{Greenstein06,Jang16,Higgins10}, stream
processing~\cite{Tatbul03,Avnur00,Raman99,Tatbul06,Cherniack03},
geo-distributed storage~\cite{Bogdanov18,Uluyol20}, and live video
analytics~\cite{Pakha18,hsu2019couper,hsieh-osdi18,jiang-sigcom18}. In
so doing, we identified four common \emph{adaptation patterns}:
\squishlist
\item \textit{Drop} eliminates data when it is no longer useful.
\item \textit{Reorder} defers lower priority data in preference to
  more important but subsequent elements.
\item \textit{Redirect} changes routing from an over-utilized resource
  to an available but possibly lower-quality one.
\item \textit{Transform} converts data from one format to
  another, typically reducing size at the cost of data fidelity.
\squishend

Each system we examined used at least one of these patterns; some
combined more than one. Importantly, all of these can be implemented
by observing at most a small, contiguous range of network messages at
the head of the current transmission sequence. Stock Envoy exposes
messages individually to stateless scripts. We expanded this interface
to allow scripts access to a single mutable ordered queue for each
(source, destination) pair. We were reluctant to widen the interface in
this way, but doing so is necessary to support reordering of messages
in applications that can benefit from it.

\subsection{\system's interface}

\system{} aims to provide an interface that is both simple
and low-overhead. This motivated a few key design
decisions. The first was to describe adaptation strategies via an
imperative Lua scripting interface. We initially explored
declarative interfaces like YAML or SQL. Unfortunately, we found it
difficult to express simple adaptations declaratively. At the same
time, these systems included significant unnecessary mechanism,
representing a potential runtime liability to the critical path of
message processing.  In addition, nearly all individual adaptation
strategies of which we are aware of have been implemented in
imperative languages, and it seemed burdensome to change models.

Second, we explicitly do not support the use of Lua libraries beyond
the standard, built-in set. This helps ensure that in-script behavior
is simple, inexpensive, and reduces the surface area for malicious
actors. If more complex functionality is necessary, it must be
provided through external callbacks, as we discuss below.

\begin{figure}[t]
  \begin{lstlisting}[style=lua,xleftmargin=13pt]
  function envoy_on_request(h)
    -- for each sink
    for queue in h:Queues():getQueue() do
      -- check the route
      route = queue:route()
      if string.find(route, "cloud") then
        -- check current bandwidth estimate
        bw = queue:getBW()
        if bw == 0 then
          -- if disconnected
          -- redirect the request to the edge
          h:redirect("edge-detector")
        elseif bw < required then
          -- if bw is too low
          -- transform the request to lower-res
          h:transform("180p")
        end
        if bw < required/2 then
          -- if bw drops well below required
          -- notify the request source
          h:notify(bw)
        end
      end end end
	\end{lstlisting}
	\caption{This simple Lua script for the traffic-monitoring application redirects requests to the edge when the network becomes disconnected, down-samples enqueued requests when bandwidth drops, and invokes a registered callback network conditions change significantly.} 
	\label{lst:yolo}
\end{figure}

\system views the world from the perspective of a single Kubernetes
pod, and how that pod should react to changes in the network.
Figure~\ref{lst:yolo} shows a sample \system{} adaptation script for a
surface street traffic monitoring application. This application uses
an ML model to detect traffic, choosing between an inexpensive one on
the nearby edge or a remote, full-fidelity version in the cloud. It
adaptively decides where to run the ML model and at what resolution.
As shown, the main abstraction exported to \system scripts is a set of
queued messages per (source, destination) pair. The scripts can iterate
over these queues (Line 3) and access various queue properties, such
as its length, route, or observed bandwidth (Line 5--8).  The queue
iterators are also used to apply in place adaptations, such as
redirecting messages to another endpoint (Line 12). Other adaptation
strategies such as dropping or reordering messages are implemented in
a similar fashion. More complex actions, such as transforming messages
(Line 16) or notifying the application of metrics (Line 21) can be
done through asynchronous callbacks.

\begin{table*}[t]
\centering
\begin{small}
	
\resizebox{\linewidth}{!}{
  \begin{tabular}{|c||p{0.12 \linewidth}|p{13cm}|p{0.1 \linewidth}|}
\hline
Context & Interface & \multicolumn{1}{|c|}{Description} &
\multicolumn{1}{|c|}{Returns} \\
\hline
\hline
\multirow{5}{*}{Queue} & 
length() & returns number of messages in a queue, useful to
approximate queuing delay.
& \multicolumn{1}{|c|}{queue length} \\

\cline{2-4} &
avgLatency() & returns weighted moving average of
end-to-end latency of messages (delta between request \&
response) 
& \multicolumn{1}{|c|}{average latency} \\

\cline{2-4} &
observedBW() 
& returns observed bandwidth allocated to the queue--the rate of the queue sending data. 
& \multicolumn{1}{|c|}{observed bandwidth} \\

\cline{2-4} &
TCPMetrics(m) 
& retrieves the TCP metrics (e.g., mean RTT) at the queue level.
& \multicolumn{1}{|c|}{TCP metric} \\

\cline{2-4} & 
messages() 
& for-loop entry to iterate over messages in the queue.
& \multicolumn{1}{|c|}{message object} \\
\hline
\hline

\multirow{12}{*}{Message} &
size() & returns the size of the message's current payload. 
& \multicolumn{1}{|c|}{size of payload} \\

\cline{2-4} &
age() 
& get the age, i.e., how long the message has been in the queue, in ms
resolution. 
& \multicolumn{1}{|c|}{age of message} \\

\cline{2-4} &
TCPMetrics(m) 
& retrieves the TCP metrics (e.g., mean RTT) at the message/request level.
& \multicolumn{1}{|c|}{TCP metric} \\

\cline{2-4} &
dst() & returns the current destination of the message. 
& \multicolumn{1}{|c|}{message destination } \\

\cline{2-4} &
header() & returns the message's header.
& \multicolumn{1}{|c|}{message header } \\

\cline{2-4} &
bytes(i, j) & returns data from {\tt i} to {\tt j} of payload of the current
message in raw binary format. 
& \multicolumn{1}{|c|}{raw payload} \\

\cline{2-4} & 
redirect(dst) & 
redirect the message to a new destination (dst).  & \\

\cline{2-4} &
transform(args)
& asynchronously transform a message's payload by forwarding to a registered endpoint.
& \\

\cline{2-4} & 
drop() 
& drops the current message from the queue. The function does not
guarantee successful operation (e.g., already transmitted in the middle of
dropping).  If successful, returns the updated queue length, otherwise, returns the old queue length.
& \multicolumn{1}{|c|}{new queue length}\\

\cline{2-4} & 
insert(msg) 
& inserts a new message {\tt msg} after the current message in the queue. If successful, returns the updated queue length, otherwise, returns the old queue length.
& \multicolumn{1}{|c|}{new queue length}\\

\cline{2-4} &
moveToFront() 
& move the message to front of the queue. &  \\

\cline{2-4} &
moveToBack() 
& move the message to end of the queue. & \\
\hline
\hline

\multirow{1}{*}{Callback} & 

notify(metrics) 
&  asynchronously send registered endpoints a metrics string. & \\

\hline
\end{tabular}

}
\end{small}
\caption{\system interface for in-network scripting. 
} 
\label{tab:bb_api}
\end{table*}

%
%
%
%

%
%
%



\noindent\textbf{Metrics exposed:} \system exports a number of network
performance metrics on which to base adaptation decisions; these are
summarized in Table~\ref{tab:bb_api}. At the
lowest level, \system exposes TCP metrics such as the congestion
window size, number of in-flight packets, and round-trip time (RTT).
\system also exposes the average end-to-end latency for messages in a
queue, which Envoy calculates using request and response arrival
times. In addition, \system provides information about how long each
messages has spent in a queue through an object-item's age property.

Network bandwidth is a crucial metric for numerous adaptation
strategies.  \system does not measure available bandwidth along a
physical link, but it calculates the observed bandwidth for messages
forwarded from a particular queue. This allows scripts to reason about
the observed bandwidth along their path of interest. For example,
scripts can detect that a path has been disconnected if its observed
bandwidth drops to zero.
 
\subsection{\system{} architecture}
\label{sec:callback}

\begin{figure}[]
	\centering
	\includegraphics[width=\columnwidth]{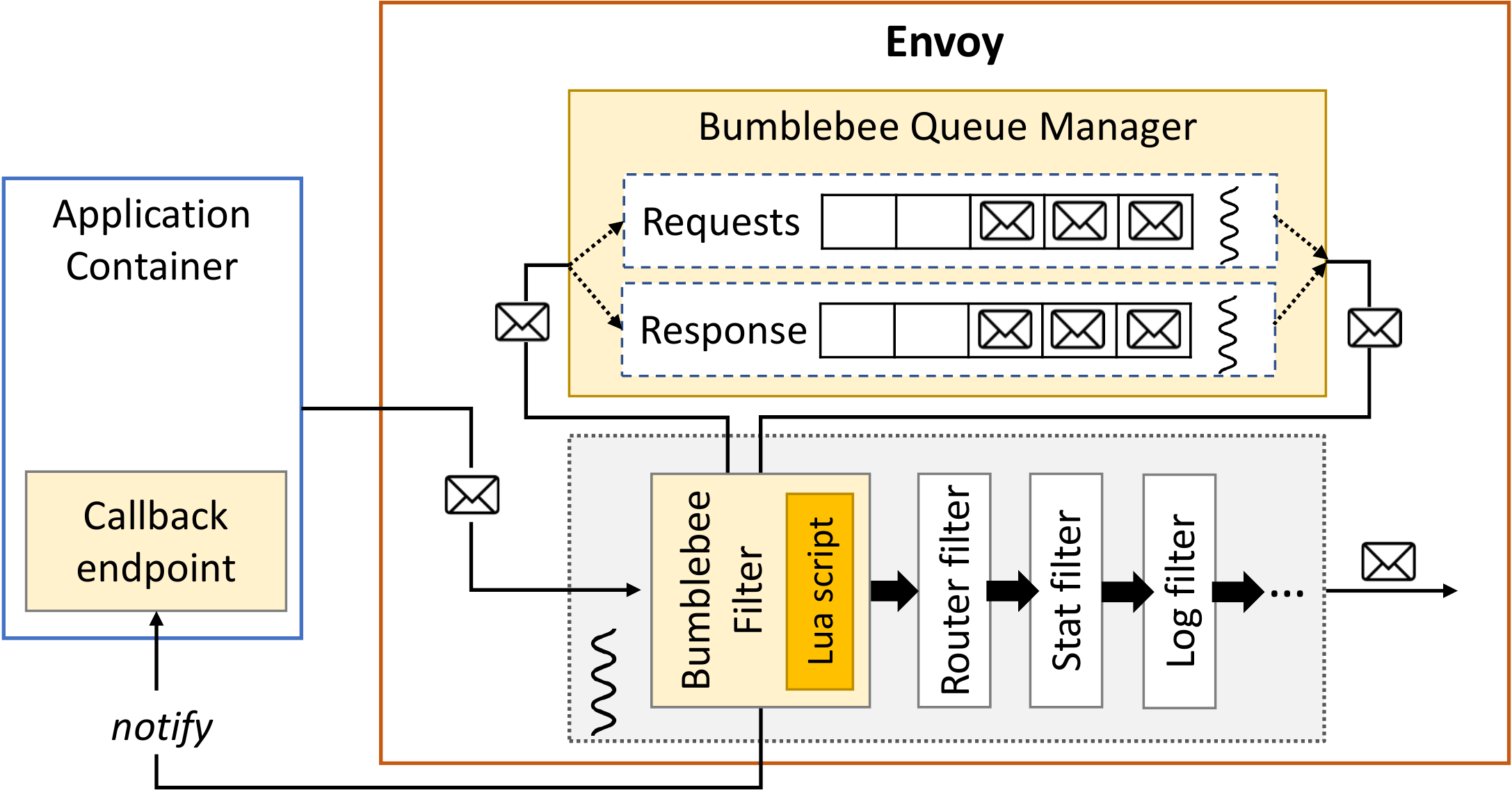}
	\caption{{\name} extends the Envoy sidecar (\name{} components marked as yellow).}
	\label{fig:envoy-with-BB}
  \vspace{-0.5cm}
\end{figure}

Figure~\ref{fig:envoy-with-BB} shows the architecture of \system{}. It
is situated within an Envoy sidecar, with user-defined adaptation
logic executed in the \system{} filter.  The rest of this section
describes the three key components of this architecture: a queue
manager that maintains message queues, an in-network scripting
facility that executes custom application logic as messages arrive,
and a callback mechanism that allows scripts to interact with other
parts of an application.

{\bf Message queues:} \system represents each (source, destination) pair
in a pod as a single, mutable queue, widening the prior interface of
stateless message processing. The latter is sufficient for common
tasks of a service mesh: load balancing, filtering, etc., but cannot
support the reordering pattern needed by many adaptive applications.
We add this abstraction with a separate queue manager that runs
concurrently with the Envoy worker pool. Worker threads pass messages
to the queue manager through the \system filter. The \system filter
buffers data until a complete network message (e.g., HTTP
request/response, Netty~\cite{netty} message) has been assembled, and
then forwards the message to the queue manager. Note that treating
messages at higher protocol layers may add some additional latency; as
we show in Section~\ref{sec:eval_overhead} this is modest.

By default the queue manager creates pairs of queues for each endpoint
pair, one in each direction, providing a handful of methods to \system
filters within the worker threads. The application can optionally
request finer-grained division of queues on a per-pod basis.  For
example, in its orchestration configuration an application may name
pods containing an object-detector running on the cloud
"cloud-object-detector.local." It can then instruct \system to create
queues in each pod's sidecar for handling requests to those specific
pods. Individual filters are invoked as messages arrive on inbound
queues, and pass messages after processing to the corresponding
outbound queue. Because operations are non-blocking, we use a simple
per-queue locking scheme to synchronize access.

In addition to events based on message arrival, the manager
periodically receives timer events that implement a token-bucket
algorithm for outbound messages. When the manager has accumulated
enough tokens, messages are forwarded to Envoy's event dispatcher.  To
minimize overhead, timers scale  the token refill rate and are only
active when the queue has pending messages.

{\bf Scripting facility:} \system applies user-defined adaptation logic
to the queues maintained by the queue manager. When a worker thread
loads a \system filter, the filter reads the appropriate script from
the orchestration configuration and launches it within a Lua runtime,
a feature natively supported by Envoy. These scripts are executed only
as messages arrive; we have chosen not to also add timer events. This
ties adaptation agility to message arrival rate; a limitation we have
not found burdensome in our limited experience so far.

{\bf External callbacks:} \system's scripting environment allows
applications to perform simple processing on enqueued messages: drop,
redirect, or reorder. However, many applications can benefit from
richer interactions between adaptive scripts and the rest of the
application. For example, an application might benefit from in-network
observation, providing early detection of bandwidth or latency
changes. Likewise, an application may want to transform message
payloads in ways that are too complex for a lightweight Lua
runtime. For example, a video streaming application may want to
downsample video chunks ahead of network constructions to prevent
head-of-line blocking. To support this kind of functionality, \system
allows scripts to make asynchronous callbacks.

There are two forms such callbacks might take. The first (and simpler)
one is used to \emph{notify} external endpoints of events within an
adaptation script. An application's orchestration configuration can
bind a list of RESTful endpoints to particular notifications within a
script, taking a string as an argument. On invocation, the Lua runtime
generates asynchronous HTTP calls with the string argument to any
endpoints listed in the orchestration configuration. This exposes
information from lower layers, and so should be used in the rare cases
when an application benefits significantly from such feedback.

The second form is used to \emph{transform} message contents. 
As with notifications, applications bind invocations to a 
RESTful endpoint through their orchestration configuration. When a
script invokes transformation on an enqueued message, 
the Lua runtime marks the entry 
asynchronously forwards the message payload to the
registered endpoint for transformation. The result is returned and
substitutes for the original message. To avoid blocking on this
operation, pending messages are marked in progress, and subsequent
messages can be sent in the interim. Transformations are typically
used for complex computations that should not take place in the
critical path of the Envoy sidecar. Transformations may optionally
access other resources--such as an external database--that 
are not possible within a sidecar limited only to the standard
libraries.

    \section{Evaluation}
\label{sec:evaluation}

To evaluate \system, we seek answers to the following questions:

\begin{smitemize}
  \item Does \system enable beneficial adaptation strategies?
  \item How difficult is writing adaptation strategies in \system?
  \item How much overhead does \system add to Envoy?
\end{smitemize}

To answer the first three questions we use our \system prototype to investigate adaptation strategies for three case-study applications. First, we use \system to help a distributed, vehicular-traffic monitoring application that adapts the quality of its object detection to changing network conditions. Second, we use \system to help a stream-processing application intelligently shed requests under bursty workloads. Finally, we use \system to help a live video-streaming service to reduce stalled playback while maintaining acceptable video resolution. To answer the last question, we run wrk2~\cite{wrk2} micro-benchmarks to measure how \system affects request latency compared to Envoy. 


We run these workloads with \system using Istio 1.4.3, Envoy 1.13.0, and clusters of virtual machines managed by Azure Kubernetes Service (AKS) 1.18.14. 

\subsection{Case study: traffic monitoring}
\label{sec:eval_yolo}


Our first case study is an emulated smart-city application that streams roadside video to machine-learning (ML) models. The ML models forward a detected vehicle's bounding box and confidence level to one or more traffic-light controllers. The controllers filter bounding boxes with confidence levels below a threshold (e.g., 50\%). The controllers use vehicle counts and locations to monitor and schedule traffic, such as reducing the time between green and red lights when road congestion is high.

Traffic monitoring is representative of many edge computing applications~\cite{noghabi-getmobile20}. The input sensors (e.g., roadside cameras) and controllers (e.g., traffic controllers) are co-located on the edge with a distributed computing pipeline between them. This pipeline must process sensor data fast enough for the controllers to respond to changes in the physical environment, and the application must operate even when network conditions are poor.

The ML pipeline can be instantiated along two paths: embedded in a resource-rich cloud environment or a lightweight edge environment. The cloud offers powerful machines and can support sophisticated and accurate ML models, whereas the edge can run a limited number of less accurate models. The application prefers results from the cloud models, and it will send frames to the cloud as long as network conditions allow it.

Detection accuracy is a key measure of fidelity for traffic monitoring. Accuracy is highest when the network allows the application to stream high-resolution frames to the cloud, but as network conditions change, the application can adapt the video stream's quality by changing frame resolution or frame rate. Low-quality streams diminish model accuracy, and high-quality streams improve accuracy. The application runs at lowest fidelity when it is disconnected from the cloud. During disconnections, the application must redirect video frames to its lightweight edge models, sacrificing accuracy for availability. 



Figure~\ref{lst:yolo} from Section~\ref{sec:design} shows a \system script that implements these trade-offs. The script iterates over an egress request queue looking for entries destined for a cloud object-detector. When bandwidth drops to zero, \system redirects requests to the edge object-detector. If bandwidth falls below a threshold, \system forwards requests to the application's transform service, which reduces frames' resolution to 180p (320x180). And if bandwidth falls well below what is required, \system notifies the sender so that it can start to send lower-resolution frames instead of relying on \system to do so.

Major cloud providers like AWS~\cite{aws-outage, aws-outage2}, Azure~\cite{azure-outage, azure-outage2}, and Google Cloud~\cite{gc-outage, gc-outage2} all suffer significant outages, and recent studies show that network conditions between the edge and cloud can be turbulent~\cite{noghabi-getmobile20,chen-imc11, yan-nsdi20}. To understand how our traffic-monitoring application behaves when edge-to-cloud connectivity is poor, we run experiments with disconnections and constricted bandwidth between the edge pods and cloud pods. Note that for our experiments we logically divide cluster nodes between the edge and cloud, but the underlying physical machines and network are entirely in Azure. Our Kubernetes cluster contains virtual machines with four vCPUs, 16\unit{GB} RAM, and a 32\unit{GB} SSD. The cluster also includes two GPU nodes with an Nvidia Tesla K80 GPU, six vCPUs, 56\unit{GB} RAM, and a 340\unit{GB} SSD. We simulate a roadside camera by streaming a highway-traffic recording from Bangkok, Thailand~\cite{traffic-video}. We use YOLOv3 as our cloud object-detection model and TinyYOLO as our edge model. Both models are trained with the COCO dataset~\cite{lin-eccv14}, which is designed to detect vehicles and passengers. 

To evaluate if the application benefits from \system, we measure the number of detected vehicles and end-to-end detection latency. The former metric influences how well the application controls traffic, and the latter influences how quickly the light controller responds to traffic changes. 

\begin{figure}[t!] \centering
\begin{subfigure}{\linewidth}
  \includegraphics[width=\columnwidth]{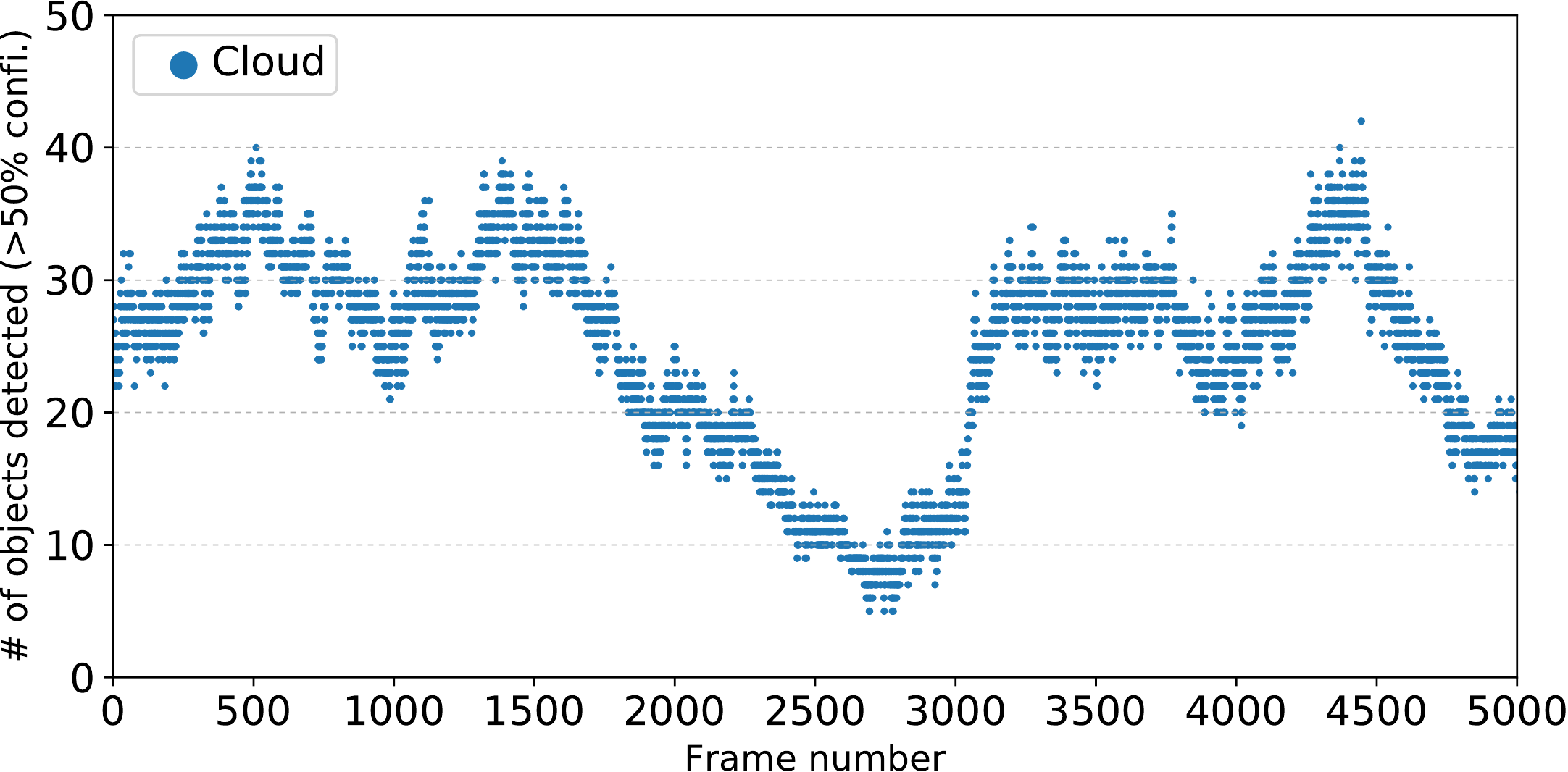}
  \caption{\bf stable baseline}
\label{fig:yolo_numobjs_cloud}
\end{subfigure}
\begin{subfigure}{\linewidth}
  \includegraphics[width=\columnwidth]{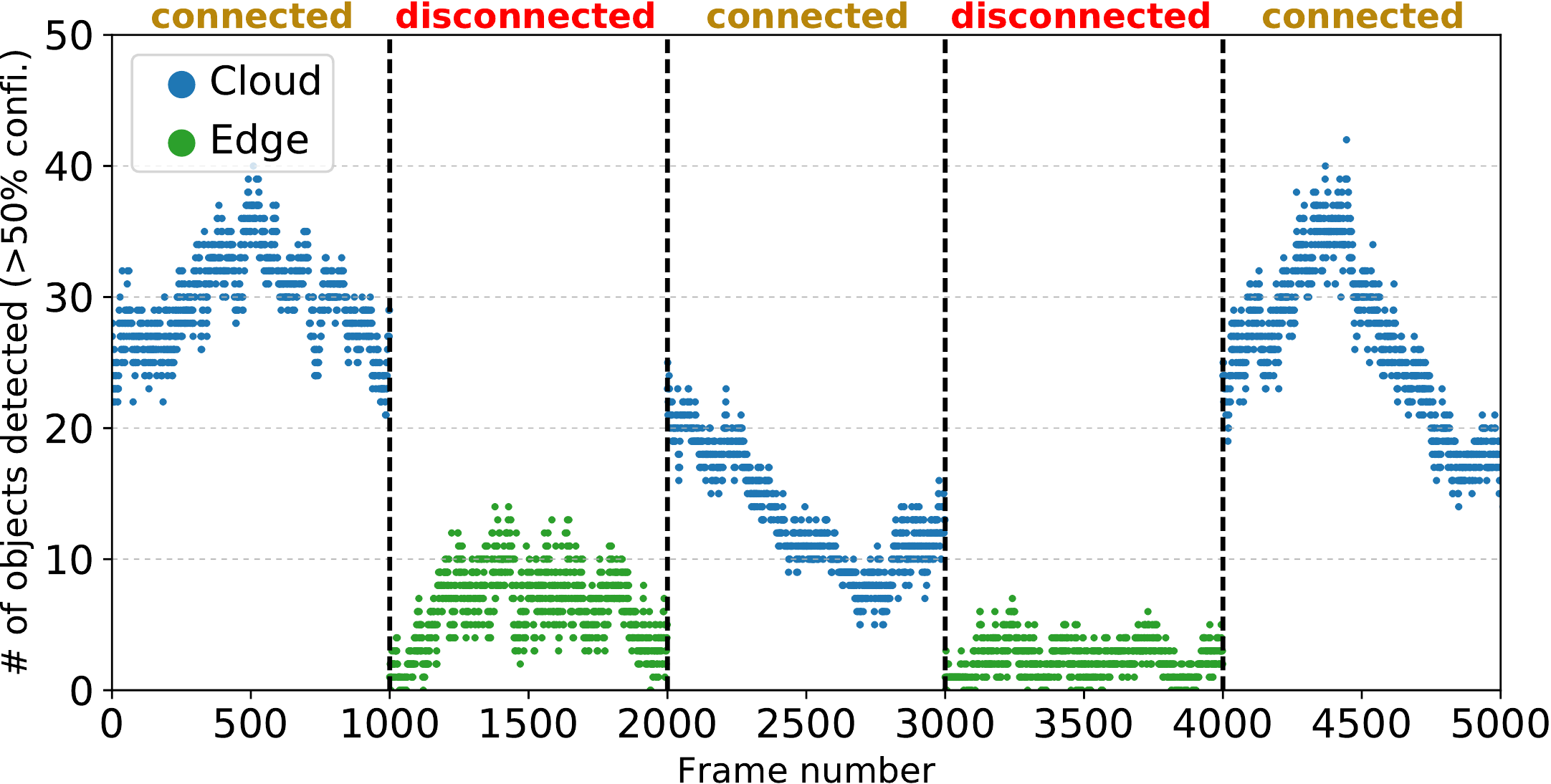}
  \caption{\bf disconnecting network}
\label{fig:yolo_numobjs_hybrid}
\end{subfigure}
\caption{
	The number of detected vehicles with greater than 50\% confidence with (a) stable network conditions and the workload running fully in the cloud, and (b) network disconnections that shift the workload to the edge using \system.
}
\end{figure}

%
%

To characterize our traffic-monitoring application without \system, we first capture the baseline object-detection accuracy of streaming 360p (640x360) video at 15\unit{fps} when fully connected to the cloud. Figure~\ref{fig:yolo_numobjs_cloud} shows the number of detected vehicles over time with a confidence threshold above 50\%. The YOLOv3 model in the cloud consistently detects between 10 and 40 vehicles. 

To simulate a disconnected edge site, we run the application under \system and partition the edge and cloud pods after 1000 and 3000 frames so that the cloud object-detector is unreachable. We heal the network between frames 2000 and 3000. Loading tensor-flow models can be slow, so \system pre-loads the edge detector at the beginning of the experiment. Figure~\ref{fig:yolo_numobjs_hybrid} shows the detected cars drop during disconnection, because the application switches to TinyYOLO on the edge. 

There is a delay between when a disconnection occurs and when \system detects the disconnection. In our experiment, five frames stall before \system detects that bandwidth is zero. Recall that our video streams at 15\unit{fps}, and so requests arrive every 67\unit{ms}. Thus, the first request sent after the disconnection experiences an approximately 350\unit{ms} of additional delay before \system redirects it to the edge. This is because four cloud-bound requests arrive after the first post-disconnection request but before \system detects the disconnection. When the sixth post-disconnection request arrives, \system has detected the disconnection and responds by redirecting all cloud-bound requests to the edge. Between disconnections, the application matches baseline detection accuracy. These results show that with \system, the application can continue to operate, albeit in a degraded mode, when the cloud is unavailable. 

\begin{figure}[t!] \centering
  \includegraphics[width=\columnwidth]{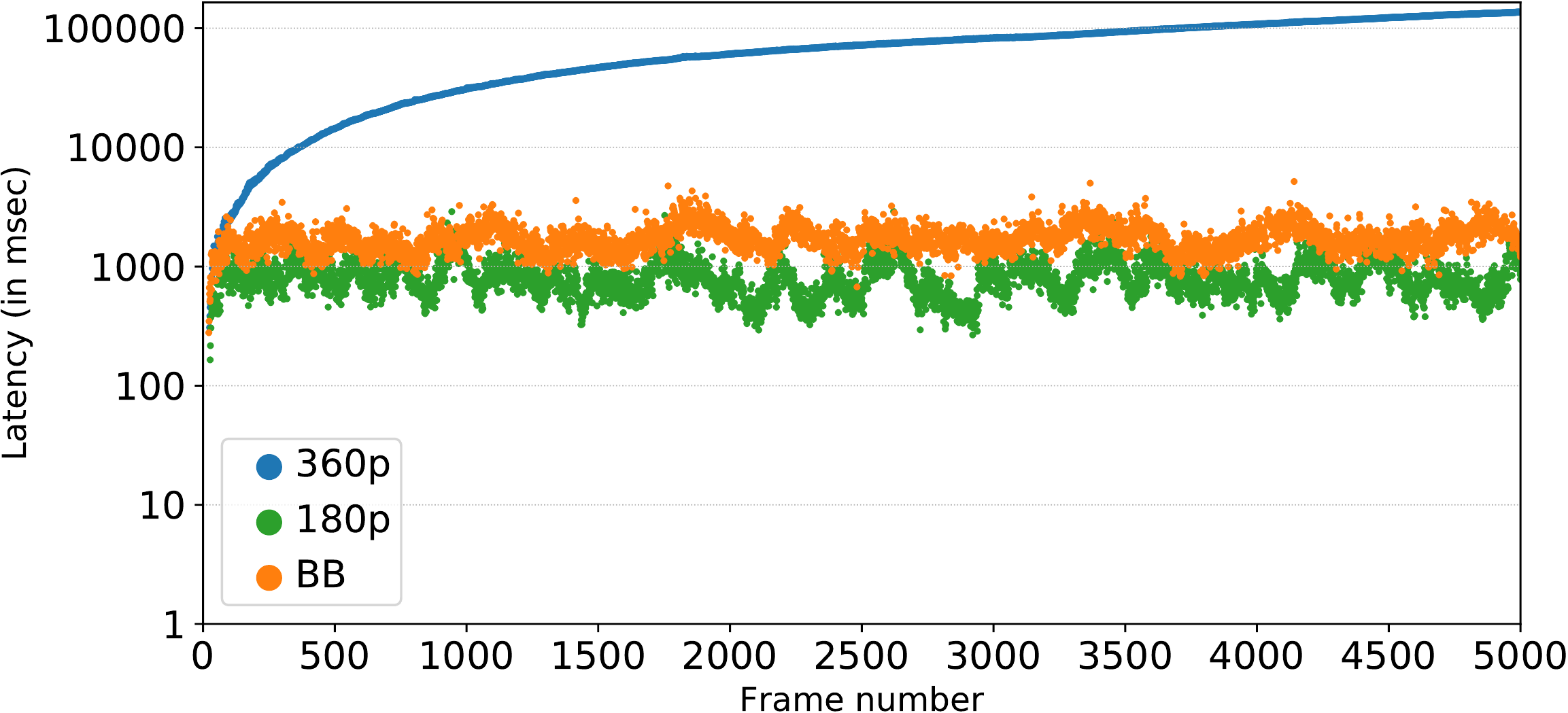}
\caption{When edge-to-cloud bandwidth is 15\unit{Mbps}, sending 360p frames leads to head-of-line blocking and exponentially increasing detection latency. Sending 180p frames reduces median latency to 815\unit{ms} with no head-of-line blocking. \system (BB) allows the application to selectively downsample frames to balance latency (median latency of 1700ms) and detection accuracy.}
\label{fig:yolo_latency}
\end{figure}

We also want to understand if the application benefits from adapting to network changes that are less dramatic than a disconnection. Recall that end-to-end latency is a critical application metric. When disconnected, the weaker edge detector processes 360p frames 33\% faster than the cloud detector using equivalent hardware. However, when bandwidth to the cloud drops, sending 360p frames can cause exponentially increasing queuing delay. To demonstrate, we restrict edge-to-cloud bandwidth to 15\unit{Mbps} and repeat the traffic-monitoring experiment twice, first sending 360p frames and then sending 180p frames. Frames are full-color, JPEG-compressed images. Figure~\ref{fig:yolo_latency} shows the results. When the application streams 360p frames (the blue line), the latency rises exponentially, but the median latency of streaming 180p frames is 772\unit{ms}. 

\begin{figure}[t!]\centering
  \includegraphics[width=\columnwidth]{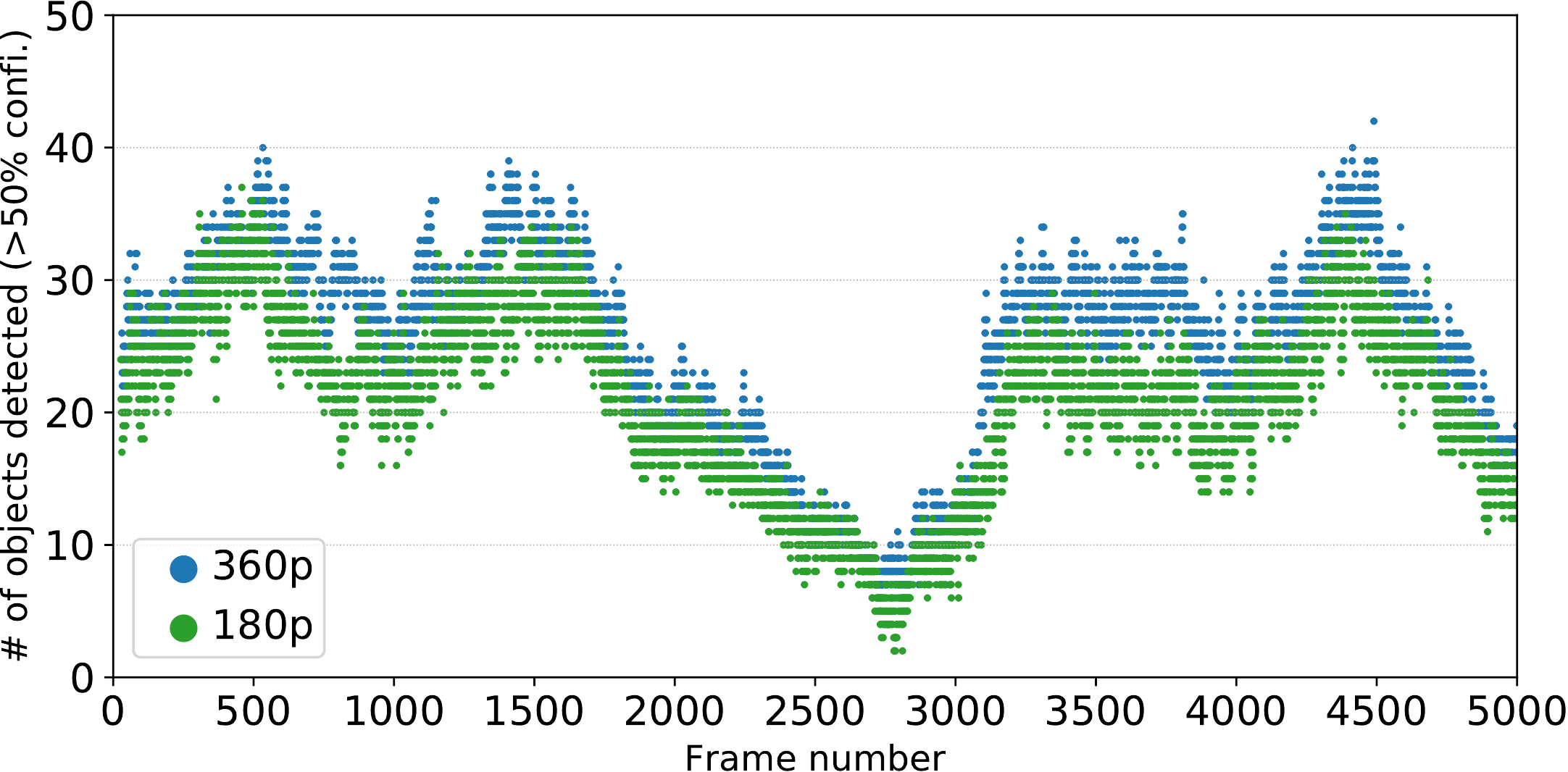}
\caption{The cloud object detector identifies more vehicles with confidence greater than 50\% in 360p frames than in 180p frames.}
\label{fig:yolo_ts_640p_vs_320p}
\end{figure}

However, lower resolution frames reduce detection accuracy. Figure~\ref{fig:yolo_ts_640p_vs_320p} shows the number of objects the application detects with confidence greater than 50\% for 360p and 180p frames. Note that the blue dots are identical to those in Figure~\ref{fig:yolo_numobjs_cloud}. 360p frames allow the cloud model to consistently detect more objects than the 180p stream, often significantly so. These results suggest that the traffic-monitoring application could benefit from selective adaptation by downsampling frames that cause queuing delay, and transmitting the remaining frames intact.
\enlargethispage{\baselineskip}

\begin{figure}[t!] \centering
  \includegraphics[width=\columnwidth]{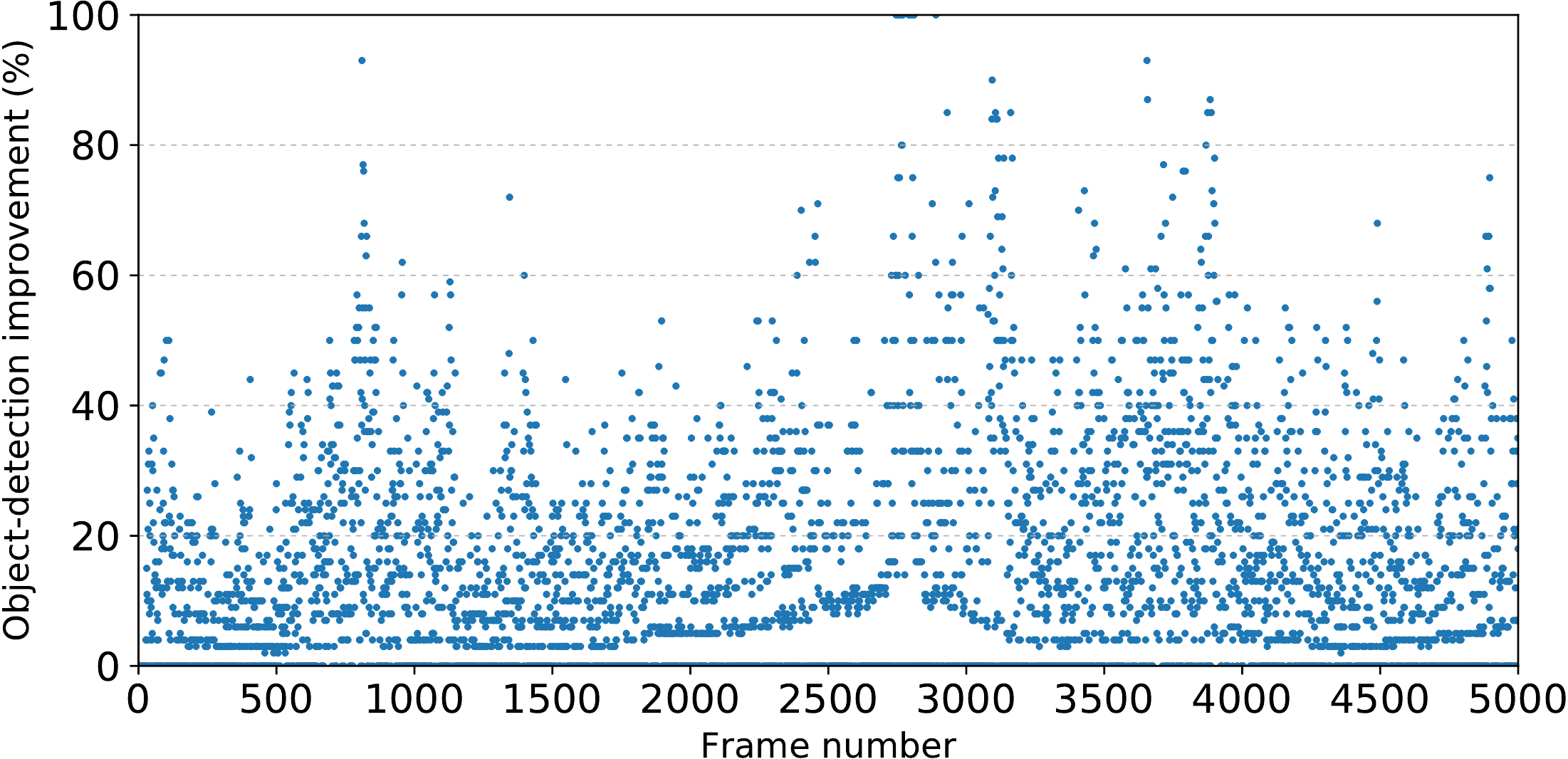}
\caption{\system enables the traffic-monitoring application to send 360p frames when possible and avoid head-of-line-blocking by selectively downsampling frames to 180p. Each blue data point represents the percentage of additional objects that the \system-enabled application detects in a frame compared to sending all 180p frames. The 3444 frames (out of 5000) are downsampled to avoid exponential queuing delays. These frames gain zero percent improvement.}
\label{fig:yolo_ts_bb_transform}
\end{figure}

To confirm our hypothesis, we repeat our limited-band\-width experiment using \system. The application sends 360p frames, and \system selectively downsamples frames that cause queuing delay. Figure~\ref{fig:yolo_ts_bb_transform} shows the percent improvement of the object detector with \system's selective downsampling enabled compared to always sending 180p frames. When \system downsamples a frame to 180p, the improvement percentage is zero. Overall, \system downsamples 3444 frames and leaves 1556 intact. Furthermore, the graph shows that selectively downsampling provides much better detection accuracy than always downsampling. Combined with the median latency of 1700ms in Figure~\ref{fig:yolo_latency}, these results show that \system allows the traffic-monitoring application to find a good balance between detection accuracy and latency using the simple script in Figure~\ref{lst:yolo}. 

To summarize, the results show that our traffic-monitoring application benefits from \system in two ways. First, the application operates when disconnected from the cloud by redirecting requests to a weaker edge object-detector. Second, when network bandwidth constricts, the application selectively downsamples frames to balance end-to-end latency and detection accuracy. We also show that the adaptation strategies responsible for these benefits can be concisely expressed by the script in Figure~\ref{lst:yolo}.

\subsection{Case-study: stream processing}
\label{sec:bb_eval_stream}

Our second case study is the Yahoo! stream-processing benchmark~\cite{Chintapalli16} that counts ad views from an input stream of ad impressions, i.e., clicks, purchases, and views. The benchmark is widely used~\cite{Venkataraman17, Hoffmann18, Mamouras2017, Zeuch19}, because it mimics in-production workloads and business logic. The first stage reads and parses impression data, the second stage filters out non-view events, and the final stage stores aggregate view counts over 10\unit{s} sliding windows. Impression counts help ad services bill customers and select the next ads to display. In the latter case, \emph{timeliness} (meeting a latency deadline) is more important than \emph{completeness} (fully processing every input), and many practitioner testimonials~\cite{Uber16, Zalando17, Pinterest19} emphasize the importance of timeliness.

By default, the Yahoo! benchmark generates emulated impressions at a constant rate, but real-world rates can be bursty. Bursts may be problematic for applications with timeliness requirements, because practitioners often statically allocate resources and must restart pipelines to scale dynamically~\cite{Venkataraman17}. Over-pro\-vision\-ing is not always possible, and unexpected bursts can rapidly increase end-to-end latency as applications fall behind processing every message.

Load-shedding~\cite{Tatbul07, Tatbul03, Xing05} is a common way to adapt to such bursts. Shedding trades completeness for timeliness by dropping less important inputs to free resources and improve the number of deadlines met. Today this adaptation strategy can only be implemented by modifying an application's internals, but \system can intelligently shed load for unmodified applications. 

To characterize how effectively \system helps the Yahoo! benchmark improve timeliness, we orchestrate the benchmark with Kubernetes by placing a containerized Apache Flink~\cite{apacheflink} worker in a pod. A worker pod can execute any stage and can pass inter-stage data within the same pod. Each Kubernetes node hosts one pod and is a virtual machine with two vCPUs and 8\unit{GB} RAM, connected by an underlying network provisioned at 1\unit{Gbps}. The benchmark polls external Kafka brokers for input events and stores results in an external Redis database. 10\unit{s} sliding windows are too coarse to properly measure the impact of bursts on timeliness, so we add a small amount of instrumentation to aggregate over 1\unit{s} sliding windows. 



\begin{figure}[t]
\begin{lstlisting}[style=lua,xleftmargin=13pt]
filt_thrd = 0.5 --- filtering threshold in sec
late_thrd = 1.0 --- lateness threshold in sec
function envoy_on_response(h)
 queues = h:Queues()
 for queue in queues:getQueue() do
   for msg in queue:messages() do 
     json = msg:json()
     if queue:avgLatency() > filt_thrd then
       event_type = json:getString("event_type")
       if event_type ~= "view" then
         msg:drop()  --- pre-emptively filter
       end 
     end

     event_time = json:getNum("event_time")
     age = h:epoch() - event_time
     if age > late_thrd then
       msg:drop()    --- drop late msgs
     end
   end end end 
\end{lstlisting}
\caption{This \system script pre-emptively filters messages and drops late messages to save inter-pod bandwidth when it detects latency in the pipeline.}
\label{lst:flink}
\end{figure}

Figure~\ref{lst:flink} shows a \system script that uses custom mess\-age-dropping logic to implement two forms of load shedding: pre-emptive filtering and dropping late messages. Recall that the benchmark filters out click and purchase events in its second stage. Under \system, if latency increases, the benchmark pre-emptively filters non-view events before the second stage (lines 8-13). The script also drops view events if they are unlikely to meet their deadline (lines 17-19). Both adaptations free resources as the script detects latency in the pipeline.




\begin{figure}[t!] \centering
  \includegraphics[width=\columnwidth]{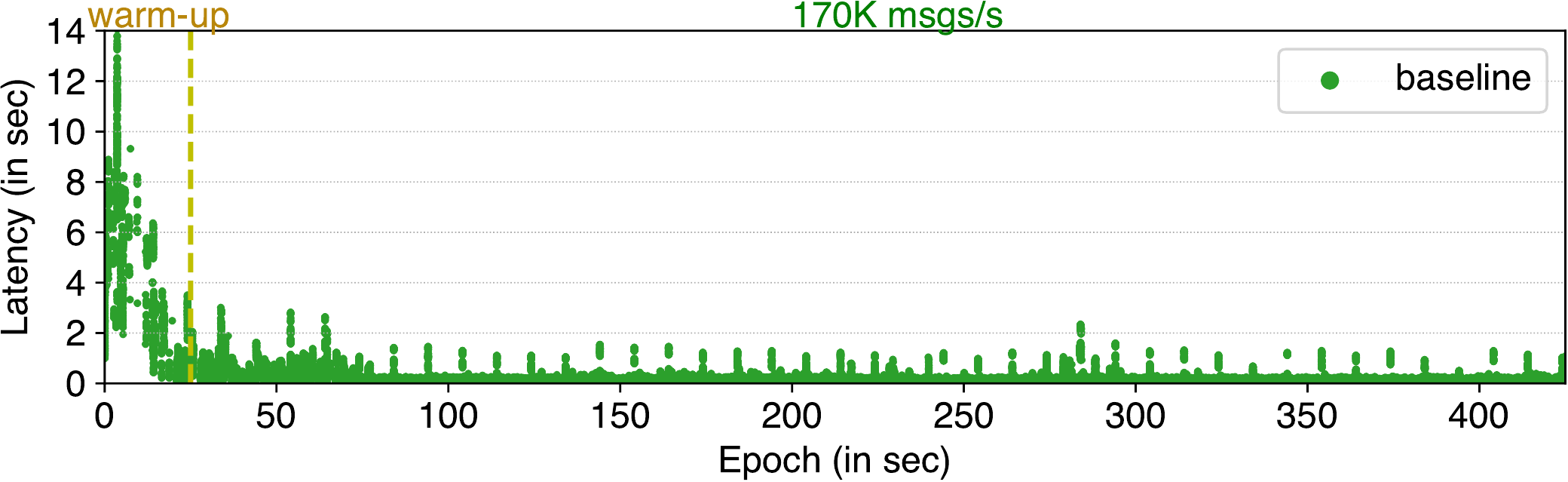}
\caption{Our stream-processing application processes input messages mostly
  under 2\unit{s} latency after a short warm-up period, when 170k input
  messages are streamed per second.
}
\label{fig:flink_baseline}
\end{figure}

Our baseline benchmark configuration runs under Kubernetes, without an
Envoy sidecar or \system. We first run the baseline configuration with
a constant, baseline load of 170k events per second. To characterize latency, we sample the end-to-end latency of the last event included in the benchmark's 1\unit{s} aggregation window. Figure~\ref{fig:flink_baseline} shows how the latency of these sampled events change over time. Latency for the first 25\unit{s} is highly variable as the benchmark warms up. After the warmup, sampled latency is largely under 2\unit{s}. This is expected since we provisioned enough compute and network resources to process every event within 2\unit{s}.



\begin{figure}[t!] \centering
  \includegraphics[width=\columnwidth]{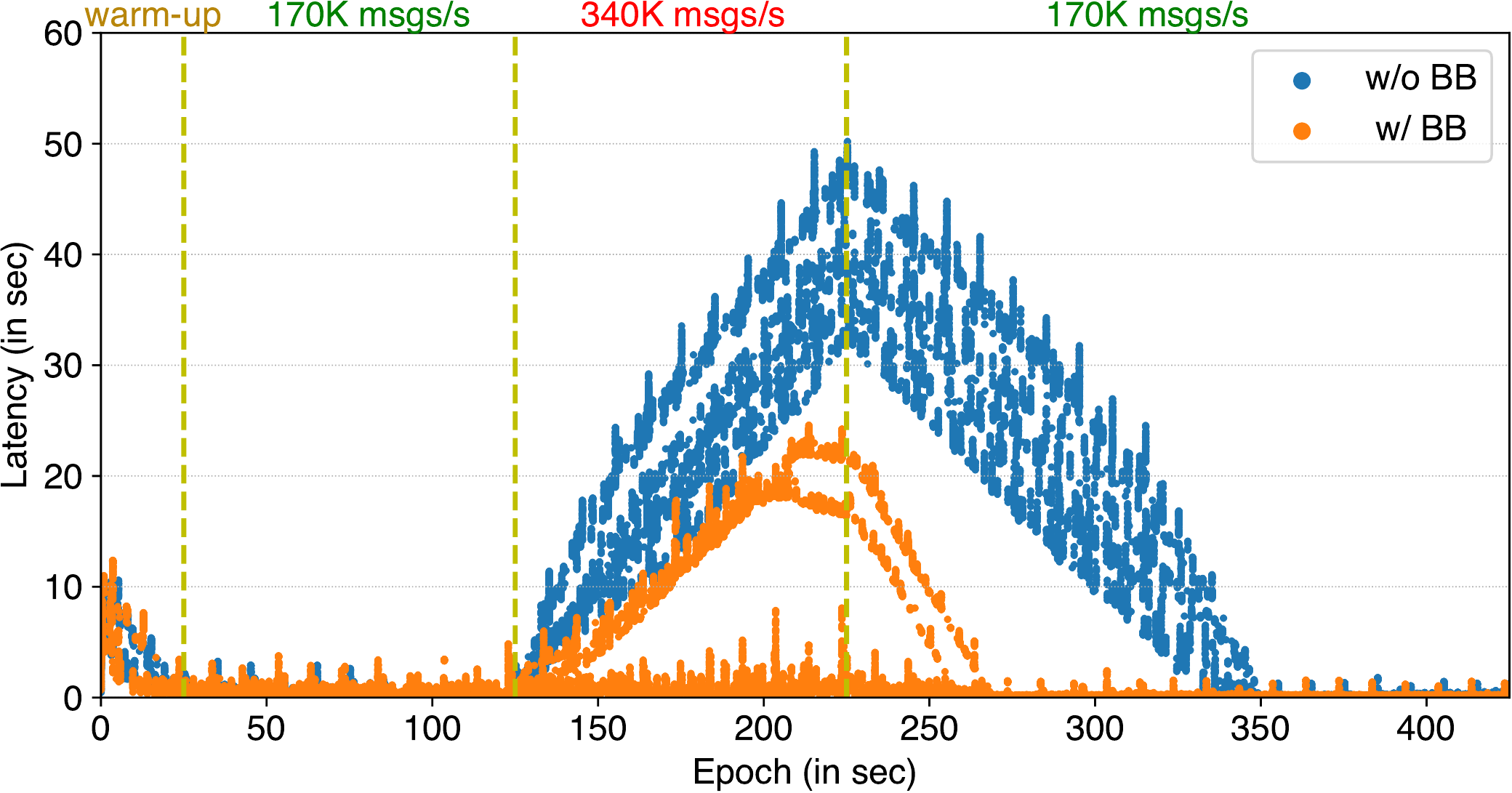}
\caption{Temporal 2x input load spikes leads the application to experience
  high latency for 1.5 times longer than the spike duration even after the
  input load returns back to the previous level. The \system-enabled
  application takes less than 30\unit{s} to bring latency back to the previous
  level.
}
\label{fig:flink_ts}
\end{figure}


We next run an experiment with variable load: first 170k events per
second for 125\unit{s}, followed by a burst of 340k events per second
for 100s, followed by a return to 170k events per second for
200s. Figure~\ref{fig:flink_ts} shows sampled latency for the baseline
benchmark (w/o BB) and the benchmark with \system (w/ BB) under
variable load. During the burst, \system drops over 29\% of all
events, and after the burst, \system drops less than 6\% of
events. Compared to the baseline, \system's custom dropping policy
significantly improves sampled latency and time to recovery. Excluding
warmup, \system allows nearly 74\% of sampled views to be processed
within 2\unit{s}, whereas the baseline benchmark allows only 44\%. In
addition, \system returns the benchmark to steady state less than
50\unit{s} after the burst ends; without \system, it returns to steady
state after 125\unit{s}.

A limitation of the current \system implementation causes the two arcs in Figure~\ref{fig:flink_ts} that peak at 20\unit{s} and 25\unit{s} sampled latency. \system intercepts only inter-pod communication, but benchmark pods contains workers for all stages. Thus, sometimes the benchmark transfers data between stage workers residing in the same pod, i.e., over local Unix sockets on which \system cannot interpose. This phenomenon is an artifact of the Yahoo! benchmark's design and would not be an issue for applications that separate each stage into a dedicated tier of pods.


\begin{figure}[t!] \centering
  \includegraphics[width=\columnwidth]{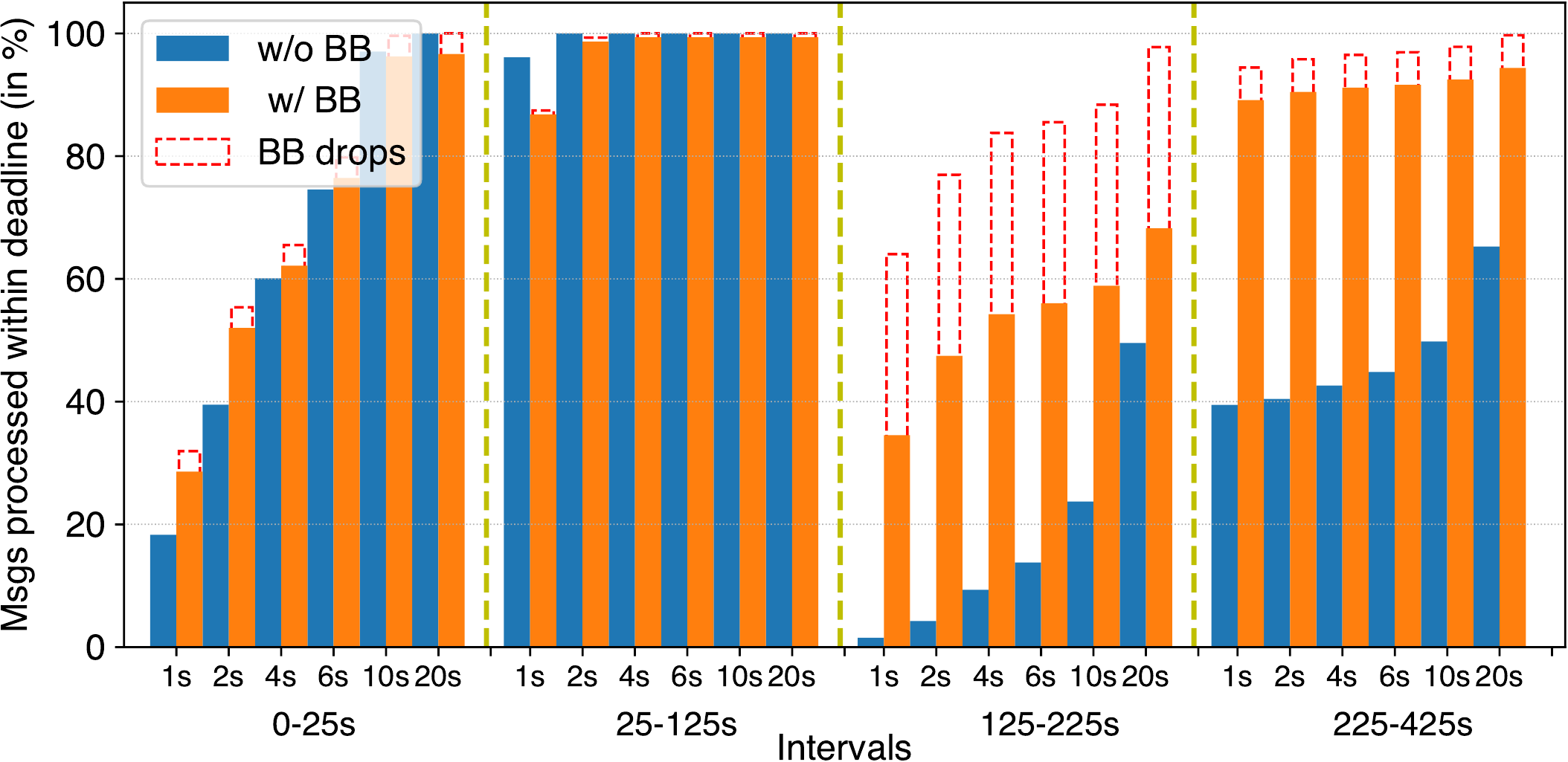}
\caption{When the input load increases above expected level that operators
  have projected and provisioned resources accordingly, the application hardly
  processes messages within a deadline. The consequence continues to stay
  longer than the ramp-up period. 
}
\label{fig:flink_bar}
\end{figure}

Figure~\ref{fig:flink_bar} highlights how \system impacts meeting deadlines of 1-20\unit{s} during the warmup, baseline-load, bursty-load, and second-baseline intervals. As expected, longer deadlines (e.g., 20\unit{s}) are met more often than shorter ones (e.g., 1\unit{s}) with and without \system. There is also little difference between the two configurations during the initial warmup and baseline intervals. However, \system provides substantial benefit during the bursty interval, allowing the benchmark to meet nearly 23x more 1\unit{s} deadlines and 37.8\% more 20\unit{s} deadlines than without \system. \system provides substantial benefits when load returns to normal, allowing the benchmark to meet over 90\% of its deadlines, regardless of length. In contrast, the baseline benchmark only meets less than 40\% of its 1\unit{s} deadlines and 65\% of its 20\unit{s} deadlines. This is due to the baseline benchmark's emphasis on completeness, and having to work through its backlog of enqueued events even after load has returned to normal.

\subsection{Case study: video streaming}
\label{sec:eval_video}


For our final case study, we evaluate an HTTP Live Streaming (HLS) service with an Nginx server and HLS.js client~\cite{hlsjs}. At runtime, the server partitions an input live stream into a rolling sequence of self-descriptive, fixed-length MPEG-TS chunks at several resolutions. When a chunk can be downloaded, the server updates an HLS manifest file to announce its availability and resolution. The HLS client is responsible for all adaptation logic and periodically polls the manifest to learn when the newest chunk is ready. After reading the manifest, the client predicts the time to download the next chunk at the available resolutions. These predictions are based on the chunks' sizes and a bandwidth estimate calculated over a sliding window of prior downloads. 

The client's competing objectives are continuous video playback and high video quality. Stalling occurs when the client's playback buffer is empty, which is far worse for the user experience than temporary drops in video quality~\cite{dobrian-sigcom11}. For example, if bandwidth drops in the middle of downloading a chunk, the client's playback buffer may drain before the transfer completes. This is common when clients react too slowly to abrupt bandwidth drops and high variability~\cite{yan-nsdi20, mao-sigcomm17}.

Prior solutions to this problem rely on either
server~\cite{mao-sigcomm17} or client~\cite{yan-nsdi20}
modifications. Modifying a server without the clean separation
provided by \system requires either building from scratch or
understanding an existing codebase and continuously merging with
external updates. Furthermore, client agnosticism is critical for open
protocols like HLS, because service providers cannot dictate which of
the many players a client may use~\cite{testing-hls}. Fortunately,
\system can transparently apply a variety of adaptation strategies to
correct an HLS client's bandwidth mis-predictions.

The \system script in Figure~\ref{lst:video} illustrates such a
strategy. The script adapts to sudden bandwidth changes faster than an
unmodified HLS.js client by only considering the most recent chunk
transfer rather than a sliding window over several transfers. Based on
this bandwidth estimate, the script chooses among available chunk
resolutions.  If the requested resolution could cause a stall (line
6-9), \system modifies the path field of the HTTP header so that it
refers to a lower-resolution chunk. However, if the client requests a
resolution that could under-utilize the available bandwidth (line 10),
\system swaps in a higher-resolution chunk path. \system's bandwidth
estimate requires Envoy modifications to monitor low-level transfer
progress or a service provider to place a middlebox between the client
and server. For the purposes of our experiments, we emulate the latter
by co-locating a proxy with the client and configuring the client to
direct its requests through the proxy. Future versions of \system will
include the necessary Envoy modifications.

\begin{figure}[t]
\begin{lstlisting}[style=lua,xleftmargin=13pt]
function envoy_on_request(h)
  hdr = h:header()
  bw = hdr:get("bw-est")
  curr, chunk = hdr:get("path")
  -- use bw estimate to choose a chunk
  pred = find_resolution(bw)
  if pred < curr then
    -- downsamples
    hdr:replace("path", pred.. "/".. chunk)
  elseif pred > curr * 2 then
    -- upsamples
    hdr:replace("path", pred.. "/".. chunk)
  end end 
\end{lstlisting}
\caption{This \system script for the video streaming application predicts
  appropriate resolution to transmit based on the most recent bandwidth
  measurement and distribution of chunk sizes. When the script disagrees with
  the client, it overwrites the {\tt path} of chunk's resolution to
  increase/decrease resolutions. Note that the script is conservative about
up-sampling to avoid potential stalls.}
\label{lst:video}
\end{figure}



We first evaluate the video-streaming service with two synthetic bandwidth changes: a sudden drop and recovery, and a gradual drop and recovery. These changes highlight the trade-offs of reacting more quickly than the HLS client's strategy. In addition, to evaluate the efficacy of \system in real-world scenarios, we analyze network-condition logs of Puffer~\cite{yan-nsdi20} clients watching live video streams. We limit our experiments to traces that cause stalls of more than 100\unit{s}, and from these traces replay estimated instantaneous bandwidth conditions. We replay the first ten minutes of each trace.
 
For all experiments we run the Nginx server under Kubernetes in a dedicated virtual machine with an Nvidia V100 GPU, 6 vCPUs, and 112\unit{GB} of RAM. For the client, the HLS.js player in the same data center as the Kubernetes cluster but in a separate virtual machine with sufficient underlying bandwidth between the two. We use Linux {\tt TC} to replay bandwidth traces at the client side, and we use the default player configuration unless noted. Each video chunk is four-seconds long.


\begin{figure*}[t]\centering
\begin{subfigure}[t]{0.33\linewidth}
  \includegraphics[width=\columnwidth]{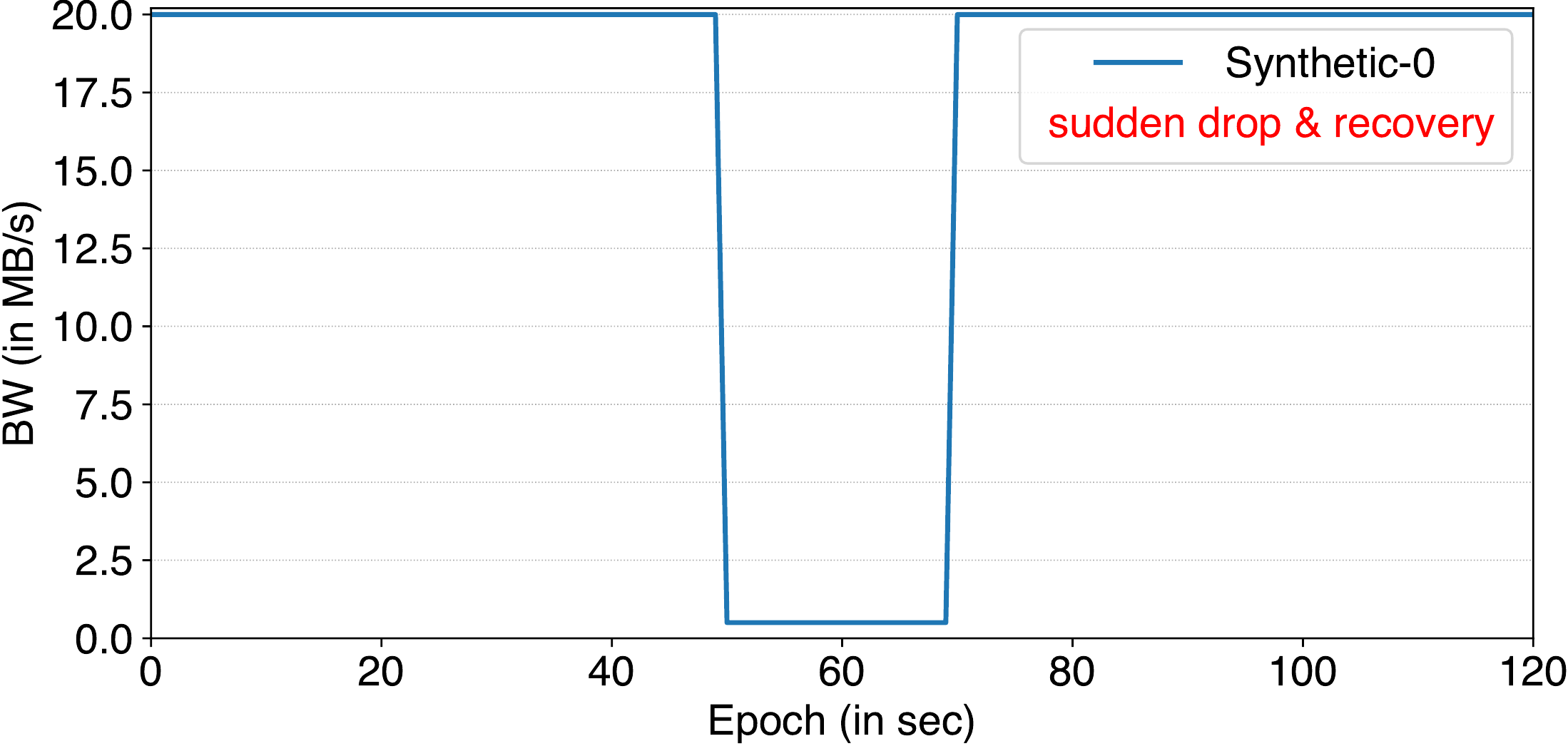}
\caption{Synthetic-0 Bandwidth Estimation}
\label{fig:video_syn0_ts_bw}
\end{subfigure}
\begin{subfigure}[t]{0.33\linewidth}
	\includegraphics[width=\columnwidth]{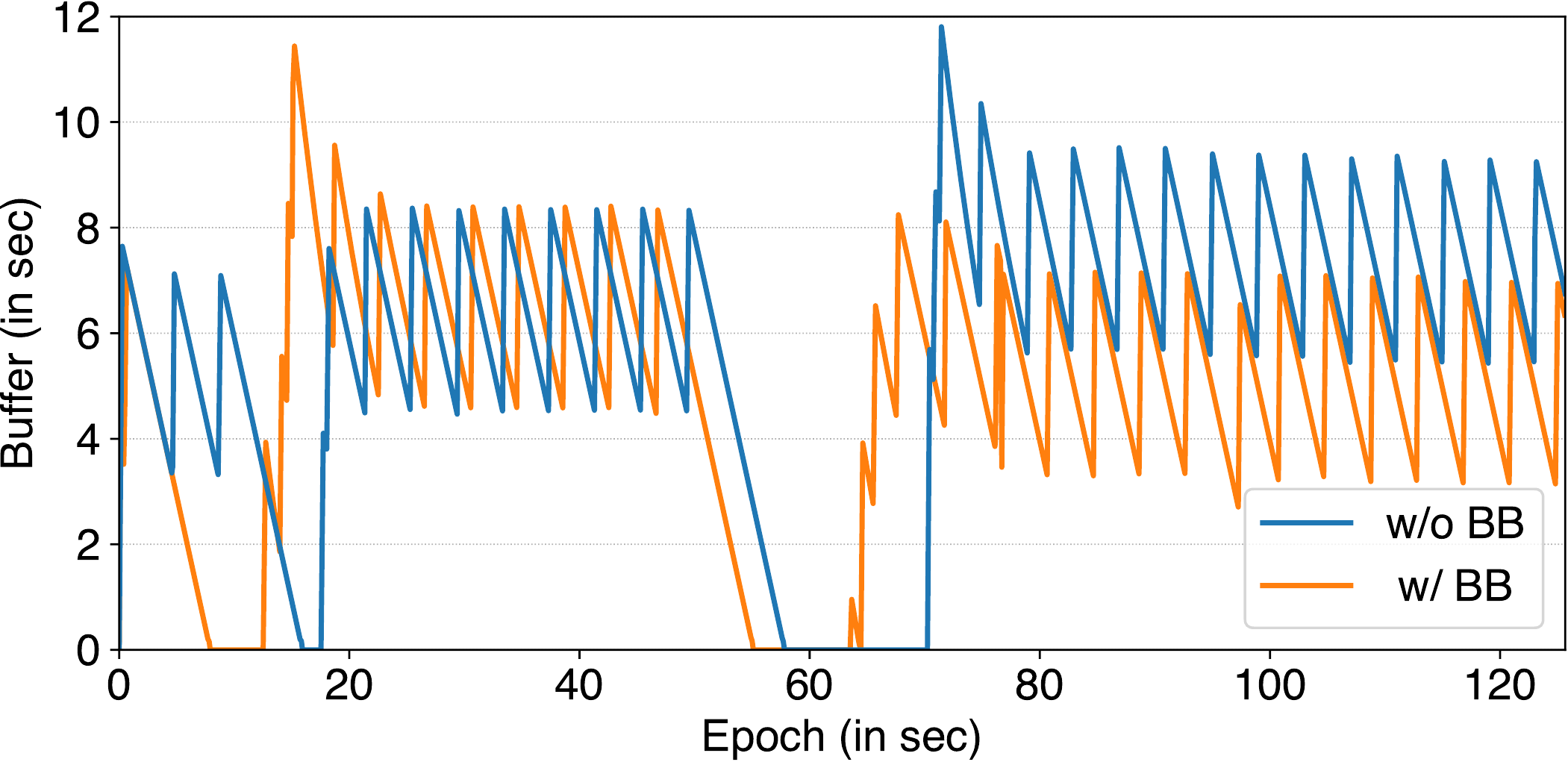}
\caption{Synthetic-0 Playback Buffer}
\label{fig:video_syn0_ts_buf}
\end{subfigure}
\begin{subfigure}[t]{0.33\linewidth}
  \includegraphics[width=\columnwidth]{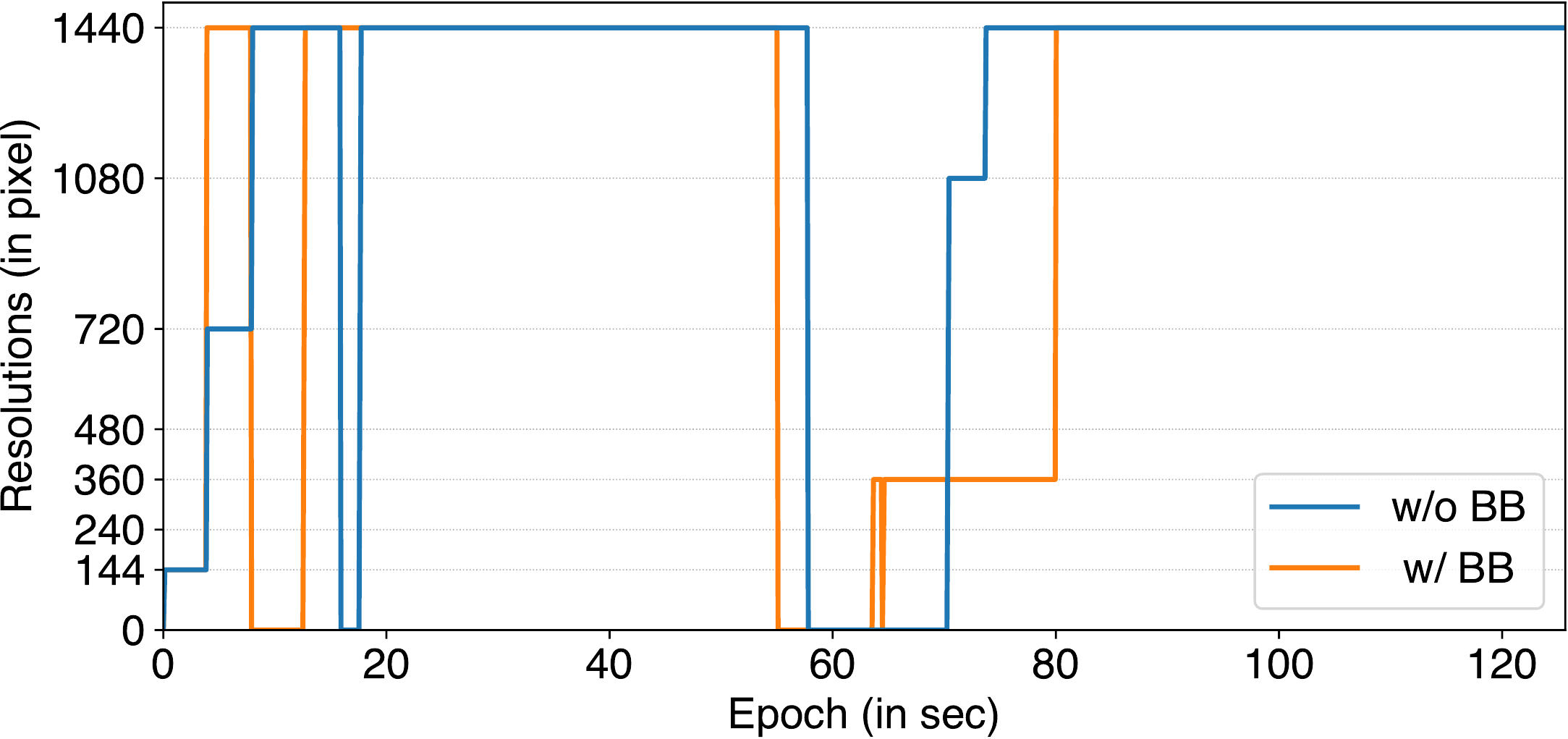}
\caption{Synthetic-0 Video Resolutions}
\label{fig:video_syn0_ts_res}
\end{subfigure}
\begin{subfigure}[t]{0.33\linewidth}
	\includegraphics[width=\columnwidth]{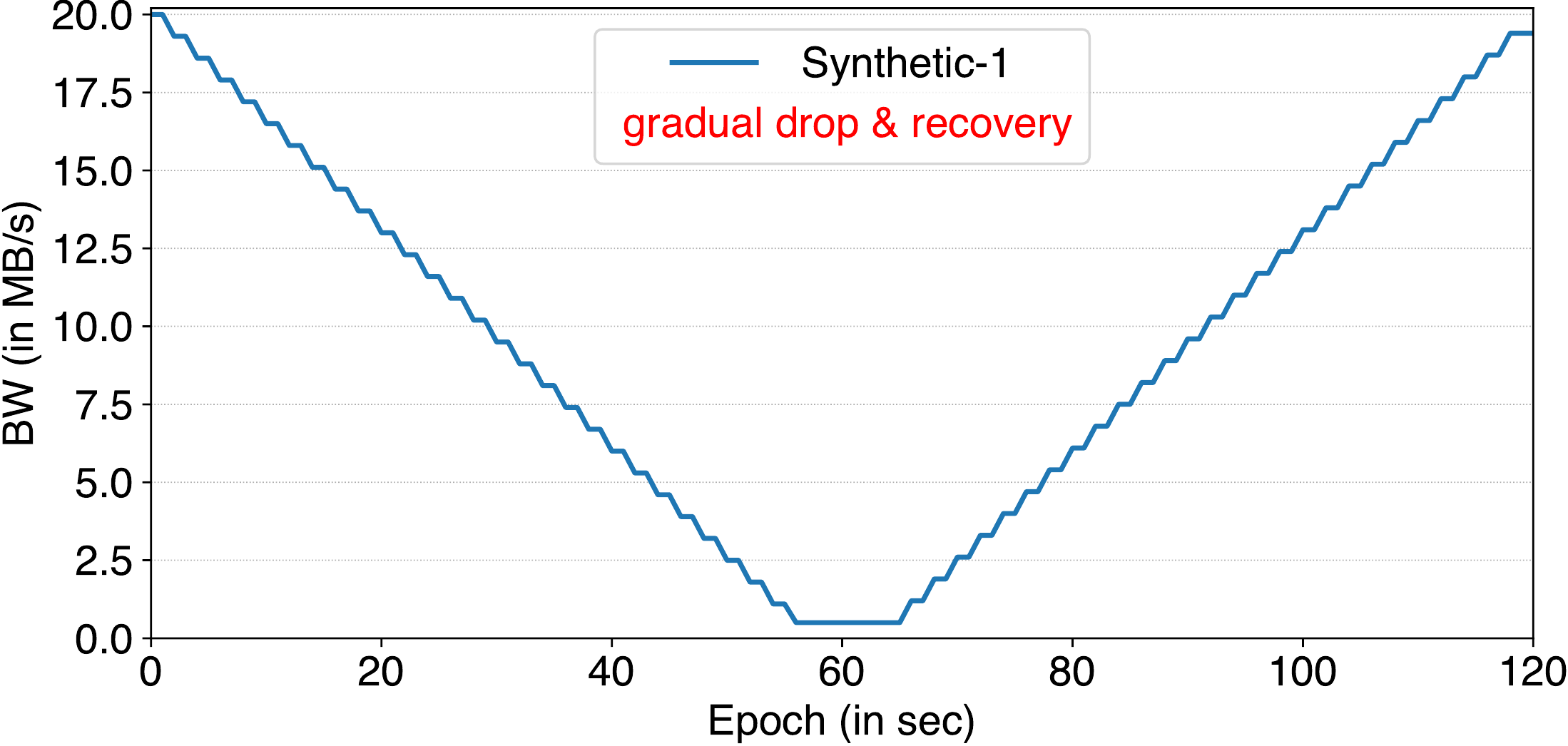}
\caption{Synthetic-1 Bandwidth Estimation}
\label{fig:video_syn1_ts_bw}
\end{subfigure}
\begin{subfigure}[t]{0.33\linewidth}
	\includegraphics[width=\columnwidth]{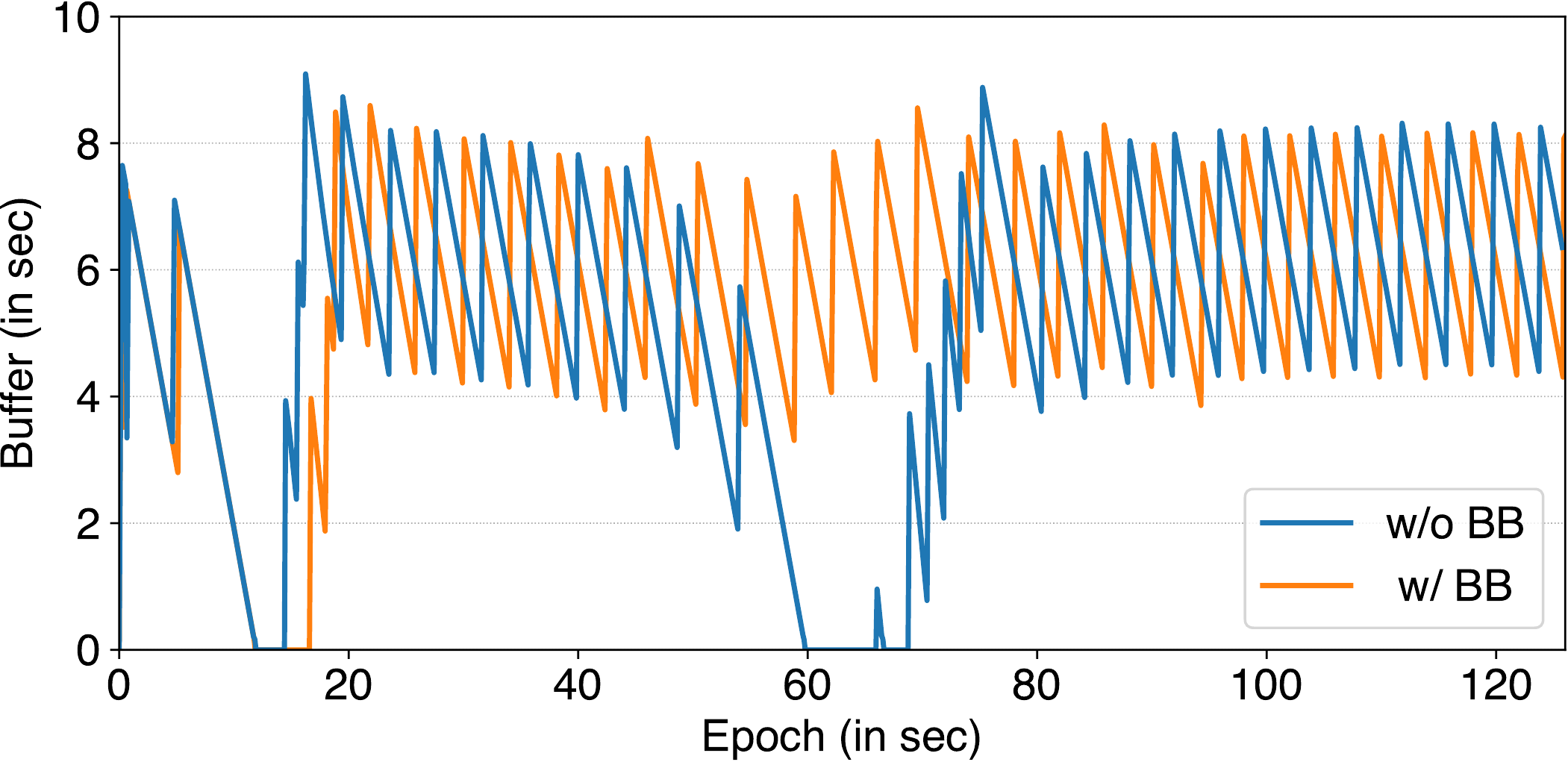}
\caption{Synthetic-1 Playback Buffer}
\label{fig:video_syn1_ts_buf}
\end{subfigure}
\begin{subfigure}[t]{0.33\linewidth}
  \includegraphics[width=\columnwidth]{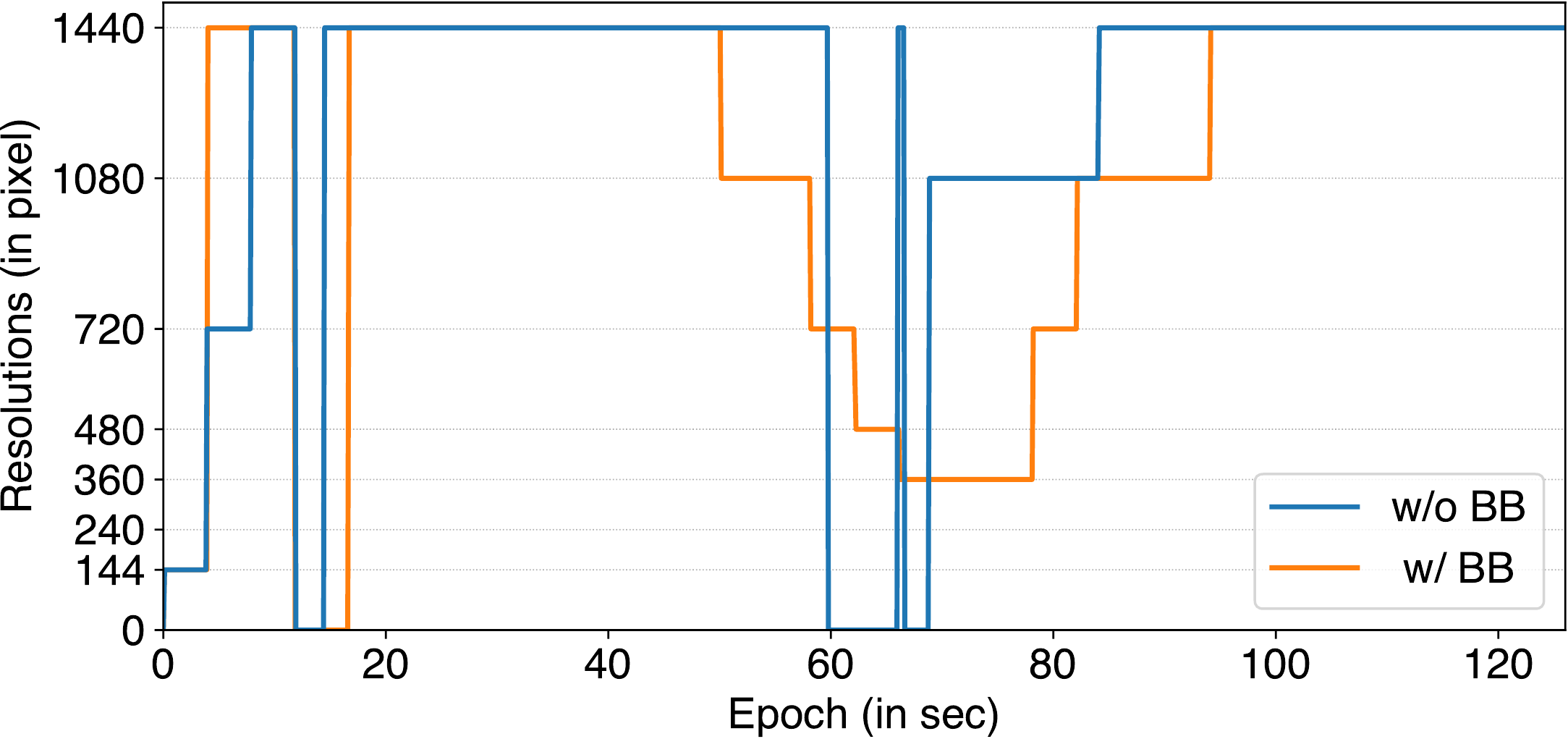}
\caption{Synthetic-1 Video Resolutions}
\label{fig:video_syn1_ts_res}
\end{subfigure}
\caption{\system helps the live video streaming application to adapt quickly
  and cautiously. Figures in the first column show the bandwidth estimates for
  both traces.  Figures in the second column show how the client's playback
  buffer changes during the trace. Figures in the last column demonstrates
  fast and agile adaptations by \system's script.
}
\label{fig:video_ts}
\end{figure*}

We characterize streaming with and without \system with two metrics: playback buffer and video resolution. The playback buffer is the seconds of video that a client can play without receiving new data from the server. When the buffer reaches zero, the video stalls. Resolution represents video quality. Buffer and resolution can be traded off. In the extremes, sending only low-resolution chunks minimizes quality but maximizes buffer, and sending only high-resolution chunks maximizes quality but minimizes buffer. Because stalls are so painful~\cite{dobrian-sigcom11}, \system wants to offer acceptable quality with minimum stalls.

Figure~\ref{fig:video_syn0_ts_bw} shows the first synthetic trace: a
sharp bandwidth drop for 20\unit{s} and fast
recovery. Figures~\ref{fig:video_syn0_ts_buf}
and~\ref{fig:video_syn0_ts_res} show the clients'
playback buffer levels and 
displayed resolutions over the course of the trace, respectively. The
client under both configurations stalls as it calibrates its bandwidth
estimates. The client under both configurations also stalls when
bandwidth drops. However, under \system the client adapts to the drop
and rebuilds its playback buffer faster than without \system. Overall,
\system reduces stalling from 13\unit{s} to 9\unit{s}, a 32\%
improvement. As Figure~\ref{fig:video_syn0_ts_res} shows this is
possible because under \system the client reduces its resolution to
360p near 65\unit{s}, whereas without \system the client continues to
download 1080p chunks.

Figure~\ref{fig:video_syn1_ts_bw} shows the second synthetic trace: gradual bandwidth decrease and recovery, each over 50\unit{s}. We hypothesized that \system would offer little benefit on this trace, anticipating that the client's default estimates would closely track the gradual changes. Surprisingly, Figure~\ref{fig:video_syn1_ts_buf} shows that \system reduces post-calibration stalling from 11\unit{s} to 5\unit{s}, a 55\% improvement. Figure~\ref{fig:video_syn1_ts_res} shows that without \system the client fails to adapt to decreasing bandwidth, continuing to fetch 1440p chunks. In comparison, \system reduces resolution in a  step-wise fashion and eliminates all stalling in the valley.

\begin{figure}[t!] \centering
  \begin{subfigure}[t]{\linewidth}
    \includegraphics[width=\columnwidth]{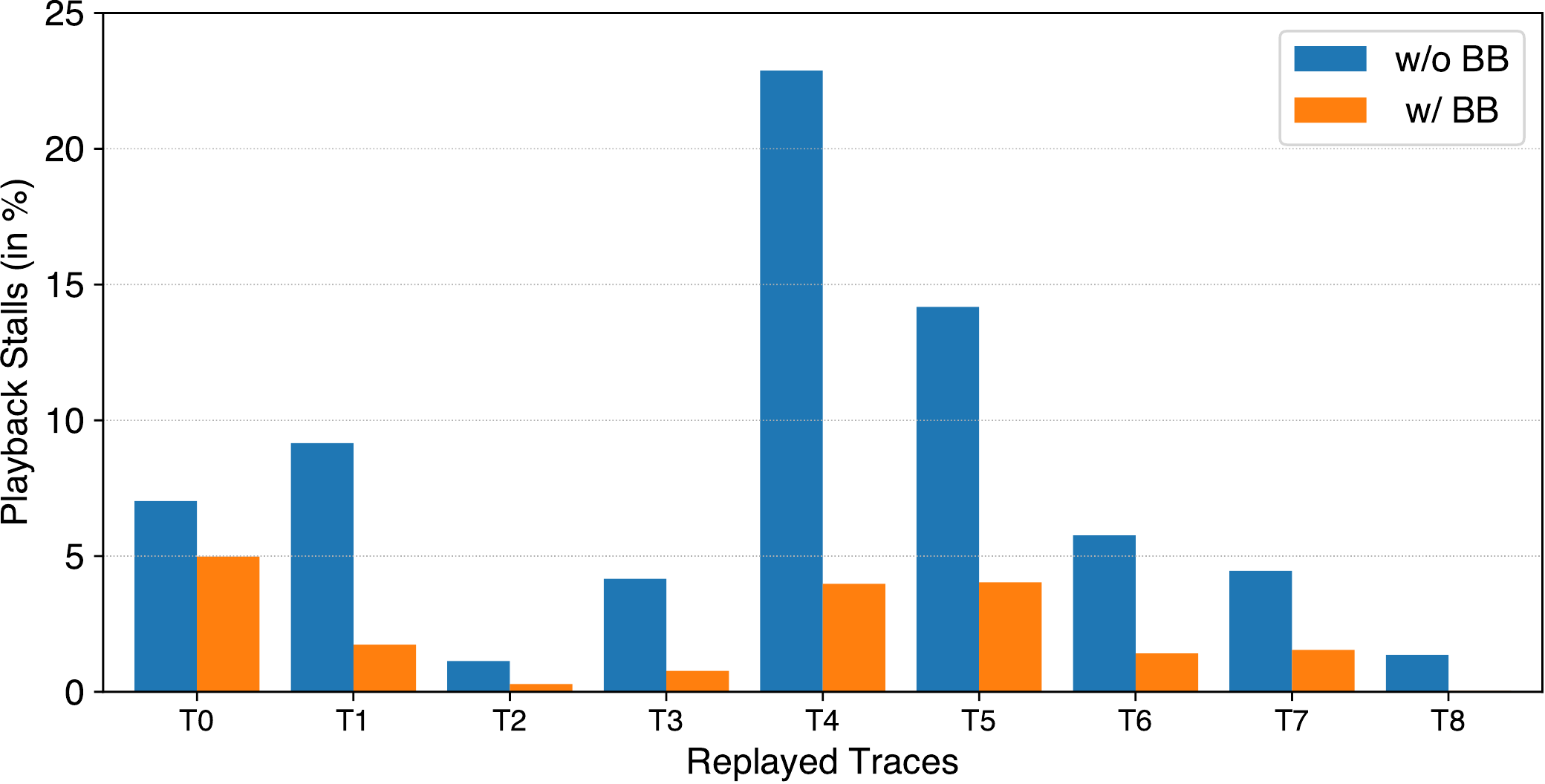}
    \caption{Across all traces, playback stalls are significantly less with
    \system than without.}
    \label{fig:video_bar_stalls}
  \end{subfigure}
  \begin{subfigure}[t]{\linewidth}
    \includegraphics[width=\columnwidth]{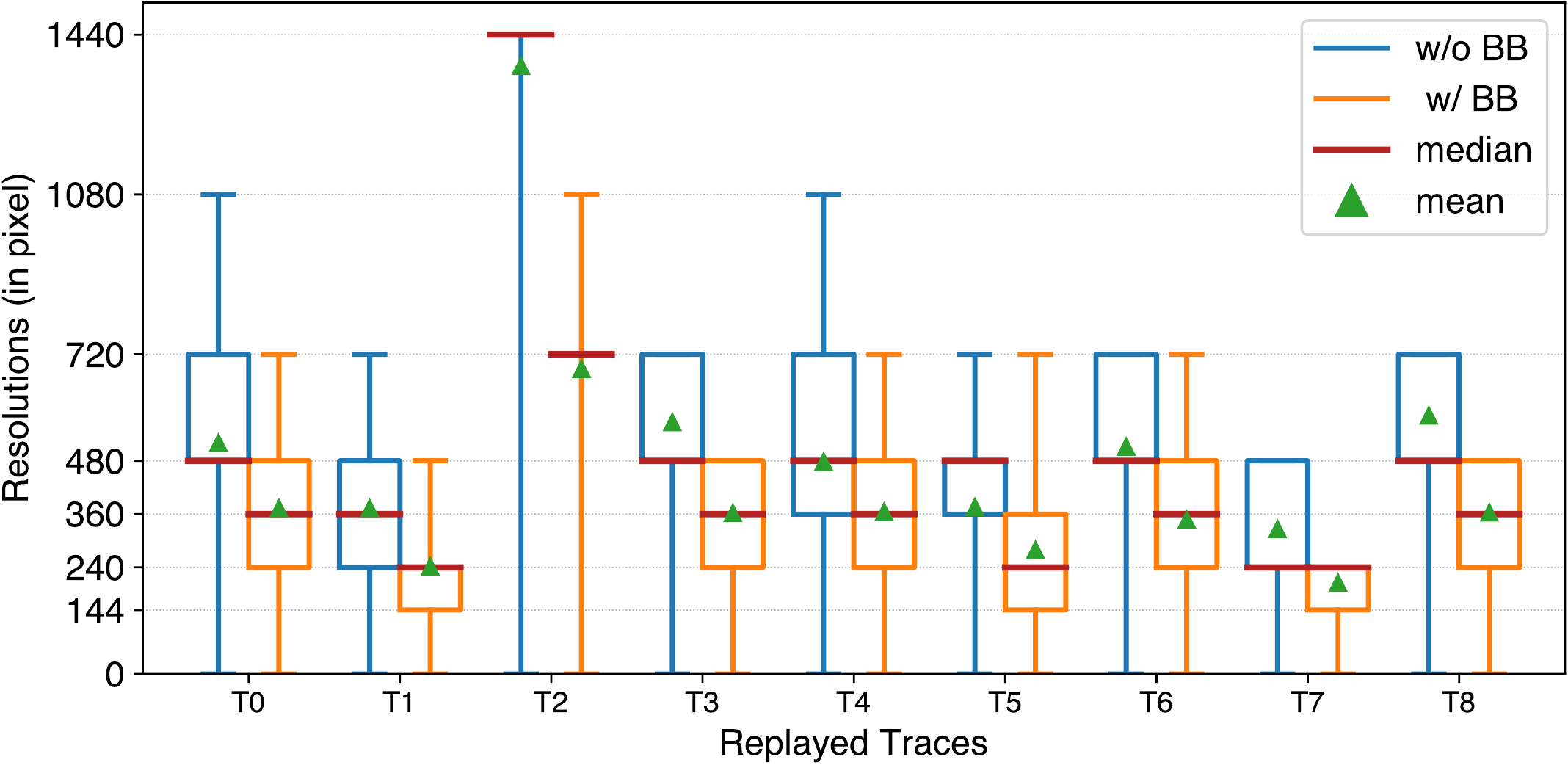}
    \caption{To eliminate stalls, \system sacrifices some video resolution.}
    \label{fig:video_box_resolutions}
  \end{subfigure}
\caption{Experiments with the nine Puffer traces with the most stalls show how
\system helps the live video streaming application to reduce stalls while
maintaining acceptable video resolution.}
\label{fig:video_puffer}
\end{figure}

We repeat the experiments with nine Puffer traces. Figure~\ref{fig:video_bar_stalls} summarizes the percentage of total stall time that a client experiences during each trace, with and without \system. The client with \system stalls at the most 5\% of the total duration, and the client without \system stalls 22\% of the time, a 77\% improvement. Figure~\ref{fig:video_box_resolutions} shows box plots of playback resolution, including mean and median. Note that in trace T2, which exhibits the least stalling without \system, the HLS.js client achieves higher resolutions than with \system albeit with some additional stalling. From the logs, we find that \system's script is too cautious about sending higher resolutions that could clog the connection during T2. This matches our expectation that quickly reacting to network changes to aggressively avoid stalling can lead to worse bandwidth utilization.


\subsection{Latency micro-benchmarks}
\label{sec:eval_overhead}


To evaluate the overhead imposed by \system compared to Envoy, we measure end-to-end latency using the HTTP benchmarking tool wrk2~\cite{wrk2}. The tool generates HTTP requests at a constant rate and outputs a latency distribution. We configure wrk2 to generate 500 requests per second with 1000 concurrent connections over five one-minutes runs. 

For our experiments, we create a client pod that runs wrk2 and assign it to a GPU node. We also create a server pod running the Nginx web server under default settings on a normal node. Both pods contain an Envoy sidecar with access to two cores. We measure the latency distribution under four client configurations: (1) Envoy without \system, (2) \system with no Lua script, (3) \system with a simple queue-iteration script, and (4) \system with a simple LIFO (Last In First Out) script. The first configuration serves as baselines for understanding \system's scripting overhead. Note that all configurations with \system move messages from the \system filter to the queue manager. 

%

We use a simple LIFO and queue-iteration scripts used in the experiments. \system's queues are internally implemented as doubly-linked lists, which makes LIFO reordering relatively inexpensive. However, iterating over the queue could be slow for two reasons. First, \system uses a per-queue locking scheme that ensures only one script can execute at a time. Second, the Lua runtime creates a new stack and object bindings each time the iteration script runs. These startup costs are drawbacks of using a scripting language instead of binary executables or bytecodes like WebAssembly. 


 To test our hypothesis, we run wrk2 five times with each client configuration. Up to the 75th percentile, the latency for all \system configurations is very close to Envoy, between 6.5\% to 12\% extra overhead, where the absolute value for the latency overhead is between $0.15ms$ and $0.35ms$. However, the cost of iterating over the queue is apparent at the very tail of the distribution. For example, at the 90th percentile, the iteration script's latency is 23\% more than Envoy's, and at the 99th percentile it is 9.5\% more. To further quantify \system's latency overhead when iterating the queue, we ran a micro-benchmark over a 10,000 length queue and called the {\tt size()} function on each request. We measured the amortized latency overhead of iterating over a single request and found this value to be negligible, an average of $0.26\mu s$. 


    \section{Related work}
\label{sec:related}

\noindent{\bf Adaptation in mobile computing:}
Resources in mobile computing are highly constrained as opposed to a data-center
environment. Prior adaptation systems~\cite{noble-sosp97, fox-asplos96}
trade application fidelity for various metrics.  Similar to
\system's callback functionality, Odyssey~\cite{noble-sosp97} creates a
collaborative adaptation solution that notifies applications to adapt their
fidelity.
On-demand distillation~\cite{fox-asplos96} performs ``on-the-fly'' adaptive
transcoding of web contents based on the client's bandwidth, similar to
\system's dynamic transformation. However, these systems do no expose control
over the enqueued data. 

Many others~\cite{balan-mobisys07, cuervo-mobisys10, chun-eurosys11} integrate
adaptation logic for better use of computation resources.  Cyber foraging~\cite{balan-mobisys07} is a runtime framework that allows developers to write and deploy complex adaptation tactics.
MAUI~\cite{cuervo-mobisys10} and CloneCloud~\cite{chun-eurosys11} partition application code, either with developer-defined annotations (MAUI) or through static analysis (CloneCloud). Then, they adaptively offload partitions between local execution (on the mobile device) to remote execution.
\system can be thought of as an extension to these systems where it can redirect the offloading traffic based on runtime variables such as network bandwidth.

{\noindent\bf Adaptation in video streaming:}
Video streaming~\cite{yan-nsdi20, qin-conext18, mao-sigocmm17,
huang-sigcomm14, spiteri-infcom16, jiang-conext12} is another domain that
employs various adaptation strategies to improve video watching experience.
A few recent works~\cite{yan-nsdi20, mao-sigcomm17} propose video streaming
servers that adaptively select the best bit-rate by using machine learning to predict the bandwidth or transmission time. Others have developed video clients to adapt to network conditions changes for fairness~\cite{jiang-conext12} and
stability~\cite{huang-sigcomm14}, to minimize
rebuffering~\cite{spiteri-infcom16}, and to handle unexpected network
outage~\cite{qin-conext18}.  While the individual solutions vary, these solutions can easily be
reimplemented in \system and leverage the low-level networking metrics and control available by \system to achieve improved performance (as shown in Section~\ref{sec:eval_video}).


{\noindent\bf Other Adaptations:}
%
Odessa~\cite{ra-mobisys11} is an adaptive runtime for partitioning and executing computer vision application remotely. The runtime balances the level of pipelining and data-parallelism to achieve low latency under variable network conditions.
Kahawai~\cite{cuervo-mobisys15} is a system for cloud gaming where clients
with modest local GPUs collaborate with powerful cloud servers to generate high-fidelity frames. Kahawai adapts to network changes by adjusting the fidelity and frame rate of frames.
Outatime~\cite{lee-mobisys15} is a speculative execution system for cloud
gaming where thin-clients send input and servers at the cloud render
speculative frames of future possible outcomes while adapting to network tail latencies.
These systems can leverage the scripting interface and in-network processing capabilities of \system to improve or simplify their adaptation strategy.


{\noindent\bf In-network Processing:}
%
The concept of in-network processing has been proposed over two decades ago where custom in-network applications are deployed at the router to provide additional functionalities, e.g., webpage caching~\cite{tennenhouse-ieee97}. Recent developments in networking hardware (e.g., smart NIC, FPGA) have led to revisiting the idea of in-network processing. Flexible programming languages such as P4~\cite{bosshart-sigcomm14} have emerged to simplify the development of in-network processing applications. As a result, many~\cite{li-osdi16, li-sosp17, gupta-sigcomm18, sultana-atc17} have explored using in-network processing for wide variety of use cases such as improving consensus protocols (NOPaxos~\cite{li-osdi16}), faster transactions (Eris~\cite{li-sosp17}), network telemetry (Sonata~\cite{gupta-sigcomm18}), or 
  improving network functionalities, e.g., DNS and NAT (Emu~\cite{sultana-atc17}).
Along the lines of these work, \system allows in-network processing of custom adaptation logic but for containerized environments such as Kubernetes.

    \raggedbottom
    \section{Conclusions}
\label{sec:conclusions}

In this paper we describe \system, a platform supporting
application-aware adaptation that is integrated with the orchestration
and service mesh mechanisms that support container-based
microservices. This is done by judiciously widening the in-network
interface in two ways. From above, applications supply simple scripts
that describe adaptive logic. From below, service mesh sidecars expose
the queue of pending messages so that these scripts can drop, reorder,
redirect, or transform those messages. Experiments with a \system
prototype demonstrate the benefits of our approach: (1) by using
\system, ML applications at the edge can utilize cloud resources
when available and operate without interruption when
disconnected, (2) \system increases the number of deadlines met between 23x and 37.8\% on the Yahoo! stream-processing benchmark, and (3) \system reduces stalled playback by 77\% during HLS video streaming under real-world network conditions.

    \pagebreak
    {
        \bibliographystyle{plain}
        \bibliography{resources/ref}
    }
    
\end{document}